\documentclass[fleqn,usenatbib]{mnras}
\usepackage[T1]{fontenc}
\usepackage{ae,aecompl}
\usepackage{graphicx}
\usepackage{amsmath}
\usepackage{amssymb}
\usepackage{url}
\graphicspath{{figure/}}

\newcommand{\kms}{\,km\,s$^{-1}$}
\newcommand{\molh}{$\text{H}_2$}
\newcommand{\Msun}{M$_\odot$}
\newcommand{\pc}[2]{pc$^{-2}$}

\defcitealias{Yung2019}{Paper I}

\defcitealias{Somerville2008}{S08}
\defcitealias{Somerville2012}{S12}
\defcitealias{Popping2014}{PST14}
\defcitealias{Somerville2015}{SPT15}
\defcitealias{Somerville2015a}{SD15}

\defcitealias{Kennicutt1998}{KS}
\defcitealias{Gnedin2011}{GK}
\defcitealias{Bigiel2008}{Big}

\title[SAM forecasts -- II. physical properties]{Semi-analytic forecasts for \textit{JWST} -- II. physical properties and scaling relations for galaxies at $\boldsymbol{z}$ = 4--10}

\author[L. Y. A. Yung et al.]{L. Y. Aaron Yung,$^{1,2}$\thanks{E-mail: yung@physics.rutgers.edu}
Rachel S. Somerville,$^{1,2}$  Gerg\"{o} Popping,$^{3}$
\newauthor Steven L. Finkelstein,$^{4}$ Harry C. Ferguson,$^{5}$ and Romeel Dav\'{e},$^{6,7,8,9}$
\\
$^{1}$ Department of Physics and Astronomy, Rutgers University, 136 Frelinghuysen Road, Piscataway, NJ 08854, USA\\
$^{2}$ Center for Computational Astrophysics, Flatiron Institute, 162 5th Ave, New York, NY 10010, USA\\
$^{3}$ Max-Planck-Institut f\"{u}r Astronomie, K\"{o}nigstuhl 17, D-69117 Heidelberg, Germany\\
$^{4}$ Department of Astronomy, The University of Texas at Austin, Austin, TX 78712, USA\\
$^{5}$ Space Telescope Science Institute, 3700 San Martin Drive, Baltimore, MD 21218\\
$^{6}$ Institute for Astronomy, University of Edinburgh, Edinburgh EH9 3HJ, UK\\
$^{7}$ University of the Western Cape, Cape Town 7535, South Africa\\
$^{8}$ South African Astronomical Observatory, Cape Town 7925, South Africa\\
$^{9}$ African Institute for Mathematical Sciences, Muizenberg, Cape Town 7945, South Africa
}

\date{Accepted XXX. Received YYY; in original form ZZZ}

\pubyear{2019}

\begin{document}
\label{firstpage}
\pagerange{\pageref{firstpage}--\pageref{lastpage}}
\maketitle

\begin{abstract}
The long-anticipated \textit{James Webb Space Telescope} (\textit{JWST}) will be able to directly detect large samples of galaxies at very high redshift. Using the well-established, computationally efficient Santa Cruz semi-analytic model, with recently implemented multiphase gas partitioning and \molh-based star formation recipes, we make predictions for a wide variety of galaxy properties for galaxy populations at $z = 4$--10. In this work, we provide forecasts for the physical properties of high-redshift galaxies and links to their photometric properties. With physical parameters calibrated only to $z\sim0$ observations, our model predictions are in good agreement with current observational constraints on stellar mass and star formation rate distribution functions up to $z \sim 8$. We also provide predictions representing wide, deep, and lensed \textit{JWST} survey configurations. We study the redshift evolution of key galaxy properties and the scaling relations among them. Taking advantage of our models' high computational efficiency, we study the impact of systematically varying the model parameters. All distribution functions and scaling relations presented in this work are available at  \url{https://www.simonsfoundation.org/semi-analytic-forecasts-for-jwst/}.
\end{abstract}

\begin{keywords}
galaxies: evolution--galaxies: formation--galaxies: high-redshifts--galaxies: star formation--galaxies: statistics
\end{keywords}

\section{Introduction}
The highly anticipated \textit{James Webb Space Telescope} (\textit{JWST}) will be equipped with extremely sensitive instruments that will be uniquely capable of detecting extremely distant faint galaxies. As postulated by the hierarchical structure formation paradigm within the $\Lambda$CDM (cosmological constant and cold dark matter) cosmological framework \citep{Blumenthal1984, White1978}, low-mass objects are expected to be fairly abundant throughout the Universe, even at high redshifts. However, their properties and abundances remain largely unconstrained since direct detections for these objects with current instruments are not possible.

In contemporary deep-field astronomical surveys, galaxy candidates at $z \gtrsim2$ have been routinely identified using the `Lyman-break' photometric selection technique \citep{Steidel1992, Steidel1993, Steidel1996}. This selection is carried out by using a set of strategically designed filters to identify the occurrence of the redshifted Lyman-limit discontinuity, caused by an intrinsic spectral break combined with absorption by intergalactic neutral hydrogen along the line of sight. Over the past decade, space-based surveys, such as the \textit{Hubble} Ultra Deep Field \citep[HUDF;][see also \citealt{Bouwens2011, Ellis2013, Oesch2013}]{Beckwith2006} and the Cosmic Assembly Near-infrared Deep Extragalactic Legacy Survey \citep[CANDELS;][]{Grogin2011, Koekemoer2011}, and ground-based surveys, such as the United Kingdom Infrared Telescope Deep Sky Survey \citep[UKIDSS;][]{Warren2007} and UltraVISTA \citep{McCracken2012}, have found nearly 2000 galaxy candidates at $z = 6 - 10$ using this technique, with faint objects reaching absolute UV magnitude $\sim -17$ at $z\sim6$. Observations can reach an even fainter detection limit in fields that are gravitationally lensed by massive foreground clusters \citep{Castellano2016, Laporte2016, Kawamata2016, Bouwens2017, Livermore2017, Lotz2017, Atek2018, Ishigaki2018}.
However, the uncertainties from magnification and foreground cluster modeling associated with these detections are relatively large \citep{Bouwens2017}.

The Near-Infrared Camera (NIRCam), one of \textit{JWST}'s on-board photometric instruments, possesses unprecedented IR sensitivity, which is expected to enable the detection of distant galaxies far below the detection limits of past or current facilities.  \textit{JWST}'s spectrometers, the Near-Infrared Spectrometer (NIRSpec) and Mid-Infrared Instrument (MIRI), will be able to provide follow-up spectroscopic studies for more luminous high redshift galaxy candidates. A sizable amount of \textit{JWST} observing time has already been allocated, dedicated to the search for high-redshift galaxies. At the beginning of its mission lifespan, \textit{JWST} is expected to carry out a number of Guaranteed Time Observation (GTO) and Early Release Science (ERS) projects that are aimed at studying galaxies and the intergalactic medium at high redshift, such as the \textit{JWST} Advanced Deep Extragalactic Survey \citep[JADES;][]{Williams2018}, the Cosmic Evolution Early Release Science survey \citep[CEERS;][]{Finkelstein2017}, Grism Lens-Amplified Survey from Space \citep[GLASS;][]{Treu2015}, and Targeting Extremely Magnified Panchromatic Lensed Arcs and Their Extended Star Formation \citep[TEMPLATES;][]{Rigby2017}. Aside from these, there are also ongoing projects that are making use of other facilities to prepare for \textit{JWST}, such as the Magellan Evolution of Galaxies Spectroscopic and Ultraviolet Reference Atlas project \citep[MEGaSaURA;][]{Rigby2018}.

A fair amount of effort has been dedicated to developing models to connect the observed, photometric properties to the underlying, inferred physical properties of these galaxies, as well as the physical processes that drive their formation. Physical properties that have been estimated directly based on observations include star formation rate (SFR), stellar mass, stellar age, metallicity, and radial size. Since the baryonic fraction and temperature of the Universe affect the properties of the first galaxies and the first stars, the physical properties of these emerging galaxy populations in the early universe are imprinted with the conditions of the cosmic environment at the time they were formed. Therefore, these galaxies also serve as indirect probes of the underlying cosmology, especially the overall matter density and the baryonic fraction. Moreover, there are collective properties that can be measured, such as the cosmic star formation rate density and global stellar mass density \citep[e.g.][]{Madau2014}.

Furthermore, combining more than one observed or inferred property can reveal a scaling relation between the two. These scaling relations among galaxy properties have been studied extensively for decades and yield fundamental insights about the physics of galaxy formation. Scaling relations may cover a large variety of properties in addition to the ones mentioned above. Some well-known examples are the Faber-Jackson relation \citep{Faber1976}, the Tully-Fisher relation \citep{Tully1977}, the Kennicutt-Schmidt relation \citep{Kennicutt1989}, the mass-metallicity (MZR) relation \citep{McClure1968, Lequeux1979, Tremonti2004, Gallazzi2005, Zahid2013}, and the stellar mass-SFR relation, sometimes called the star formation main sequence \citep[e.g.][]{Brinchmann2000, Brinchmann2004, Noeske2007, Wuyts2011}. We may also consider scaling relations between quantities predicted by theory that cannot be directly measured from observations, such as the stellar mass vs. halo mass relation or galaxy size vs. halo size relation (\citealt{Moster2010, Moster2013}, \citealt*{Behroozi2013a}, \citealt{Somerville2018}). These relations are collectively affected by many physical processes, both local (ISM scale) and global (galaxy scale), which provide extremely important insights regarding the formation and assembly histories of these objects. On the other hand, the intrinsic scatters in these relations also hint at whether there are important higher order parameters (i.e. whether the scaling relation actually sits in a higher dimensional space, such as fundamental plane relations). Next-generation observing facilities will constrain the evolution of these scaling relations over cosmic time.

One of the main science goals of \textit{JWST} is to constrain the nature of the sources that reionized the Universe. Recent studies have shown that high-redshift low-mass galaxies could have been the major source of the ionizing photons that reionized the Universe \citep{Kuhlen2012, Anderson2017, Finkelstein2019}. However, there has historically been tension between the efficient SF in low mass halos needed to reionize the Universe early enough to satisfy observational constraints, and the inefficient SF observed in low-mass galaxies today \citep{Madau2008, Finkelstein2015, Robertson2015}.

Models and theoretical simulations set within the framework of cosmological structure formation are a powerful tool for creating forecasts for future observations. Inevitably, there are trade-offs between the breadth of physical processes that can be included, the accuracy with which these processes can be modeled, and computational limitations. A variety of different methods have been developed and employed to make predictions for galaxy populations that lie outside of the scope of current observations. These include (sub)-halo abundance matching models, (semi-)empirical models, semi-analytic models, and numerical methods. There are advantages and disadvantages to each method, and on the whole they represent a complementary toolkit in this challenging landscape.

There has been a long standing history of using numerical methods to carry out \emph{a priori} galaxy formation simulations, in which one attempts as much as possible to simulate the main physical processes explicitly. However, in practice, since these physical processes that influence galaxy formation operate over an extremely broad range of spatial and temporal scales, one must be selective about the physical prescriptions and resolution. Inevitably, in large scale simulations, some `sub-grid' recipes are needed to represent processes occurring at scales well below the physical resolution (see \citealt{Somerville2015a} for a detailed discussion). In addition, modeling galaxy formation in a cosmological context is very challenging, given the tension between simulated volume and mass and spatial resolution.  Moreover, due to computational limitations, numerical simulations focussed on the high-redshift Universe are often halted at some intermediate ($z \gtrsim 6$) redshift (e.g. \citealt{OShea2015, Wilkins2017}; \citealt*{Jaacks2018}) and thus cannot be validated with low-redshift observational constraints. Zoom-in simulations are designed to allow high-resolution simulations to be conducted within a proper cosmological context, and are carried out by simulating a large cosmological volume at a coarse resolution and selecting and `resimulating' a desired smaller region with higher resolution and additional prescriptions for baryonic physics \citep[e.g.][]{Hopkins2014, Hopkins2018, OShea2015}. However, these techniques do not by themselves yield predictions for statistical properties of populations, as the re-simulated halos are in general not representative of the underlying full cosmological distribution.

At the other end of the spectrum, purely empirical models extrapolate observed galaxy populations to higher redshifts and/or lower luminosities with no theoretical underpinning and no cosmological framework  \citep[e.g.][]{Kuhlen2012, Williams2018}. Halo abundance matching models (and variants sometimes called semi-empirical models) attempt to derive relationships between dark matter halo properties and galaxy observables using lower redshift observations, then use the predicted evolution of halo properties from a cosmological model to make predictions for galaxy properties (\citealt{Trenti2010}, \citealt*{Trenti2015}; \citealt{Tacchella2013, Behroozi2015, Behroozi2018}; \citealt*{Mason2015}; \citealt*{Moster2018}; \citealt{Tacchella2018, Wechsler2018}). Naturally, this requires assumptions about how the galaxy-halo relationship evolves in this unexplored territory, but these approaches are highly flexible and computationally efficient.

The semi-analytic modeling approach provides an attractive `middle way' for modeling large
populations of galaxies and exploring a large dynamical range in halo mass and environment, and has been used for decades to make predictions for the properties of high-redshift galaxies (e.g. \citealt*{Somerville2001}; \citealt{Somerville2015, Henriques2015, Lacey2016, Poole2016}; \citealt*{Rodrigues2017}). This approach is built within the framework of cosmological structure formation, and adopts parameterized phenomenogical recipes for physical processes such as cosmological accretion and cooling, star formation and black hole growth, and feedback from massive stars, supernovae, and AGN. The parameters in these recipes are typically calibrated to reproduce key observational relationships for nearby galaxies. An advantage over the semi-empirical approach is that the parameters represent \emph{physical quantities} that can frequently be constrained via observations, and compared with similar quantities in fully numerical hydrodynamical simulations. At the same time, the approach is still highly computationally efficient, allowing exploration of parameter space and different physical recipes.

In this series of \textit{Semi-analytic forecasts for JWST} papers, we make predictions for galaxy populations at $z = 4$--10 with forecasts tailored specifically for upcoming \textit{JWST} observations, and we investigate how uncertainties in the physical processes may affect the global properties of these galaxies. In \citet[hereafter \citetalias{Yung2019}]{Yung2019}, we presented the rest-frame UV luminosity functions and a series of one-point distribution functions for observer-frame IR magnitudes convolved with \textit{JWST} NIRCam broadband filters. In this companion work (Paper II), we present the physical properties and scaling relations for the same set of predicted galaxies. In an upcoming Paper III (Yung et al. in prep), we will present predictions for the production rate of ionizing photons based on stellar population synthesis models and the star formation and metal enrichment histories of high-redshift galaxies predicted by our models. And in Paper IV (Yung et al. in prep), we combine our SAM with an analytic reionization model to create a physically motivated, source-driven pipeline to efficiently explore the implications of the predicted galaxy populations for cosmic reionization. All results presented in the paper series will be made available at \url{https://www.simonsfoundation.org/semi-analytic-forecasts-for-jwst/}. We plan on making full object catalogs available after the publication of the full series of papers.

The key components of this work are summarized as follows: the semi-analytic framework used in this work is summarized briefly in \S\ref{sec:sam}. We present the physical properties and scaling relations for these galaxy populations in \S\ref{sec:results}--\ref{sec:scaling_z}. We then discuss our findings in \S\ref{sec:discussion}, and summary and conclusions follow in \S\ref{sec:snc}.

\section{The semi-analytic framework}
\label{sec:sam}
The Santa Cruz semi-analytic model used in this work is slightly modified from the one outlined in \citet*[hereafter \citetalias{Somerville2015}]{Somerville2015}. We have implemented the \citet*{Okamoto2008} photoionization feedback recipe and updated the cosmological parameters to be consistent with the ones reported by the \citeauthor{Planck2016} in 2015. The model components that are essential to this work have been concisely summarized in \citetalias{Yung2019} and we refer the reader to the following works for full details of the modeling framework: \citet{Somerville1999}; \citet*{Somerville2001}; \citet{Somerville2008, Somerville2012}; \citet*[hereafter \citetalias{Popping2014}]{Popping2014} and \citetalias{Somerville2015}. The cosmological parameters adopted in this work are the following: $\Omega_\text{m} = 0.308$, $\Omega_\Lambda = 0.692$, $H_0 = 67.8$\kms Mpc$^{-1}$, $\sigma_8 = 0.831$, and $n_s = 0.9665$.

Dark matter halo merger histories, also commonly referred to as merger trees, are the backbone of our semi-analytic models for galaxy formation. In order to efficiently sample halos over a wide mass range, including halos from close to the atomic cooling limit to the rarest, massive objects expected to be found in high-redshift surveys, we adopted a merger tree algorithm based on the Extended Press-Schechter (EPS) formalism \citep{Press1974, Lacey1993}. These semi-analytic merger histories have been shown to be qualitatively similar to the ones extracted from $N$-body simulations \citep{Somerville1999a, Somerville2008, Zhang2008, Jiang2014}. At each output redshift, we set up a grid of root halos spanning the range in virial velocity $V_\text{vir} \approx 20$--500 \kms, and assign their expected volume-averaged abundances based on the halo mass function from the Bolshoi-Planck simulation from the MultiDark suite \citep{Klypin2016} with fitting functions provided in \citet{Rodriguez-Puebla2016}. For each root halo in the grid, one hundred Monte Carlo realizations of the merger histories are generated, each traced down to progenitors of a minimum resolution mass of either $M_\text{res} \sim 10^{10}$\Msun\ or 1/100th of the root halo mass, whichever is smaller.

As implemented in the latest iteration of the model \citepalias{Popping2014, Somerville2015}, the disk component of each galaxy is divided into annuli and the cold gas content in each annulus is partitioned into an atomic (\ion{H}{i}), ionized (\ion{H}{ii}), and molecular (\molh) component. Among the models implemented and tested in \citetalias{Popping2014} and \citetalias{Somerville2015}, the metallicity-based, UV-background-dependent recipe, which is based on simulations by  \citet[hereafter \citetalias{Gnedin2011}]{Gnedin2011}, yields the best results and hence was adopted as the fiducial model for multiphase gas partitioning. With the estimated surface density of molecular hydrogen ($\Sigma_\text{\molh}$), observationally motivated, empirical \molh-based SF relations are used to model the surface density of SFR ($\Sigma_\text{SFR}$) by \citet[hereafter \citetalias{Bigiel2008}; see also \citealt{Wong2002}, \citealt{Bigiel2011}, \citealt{Leroy2011}]{Bigiel2008}. Recently, evidence from both theory and observation suggests that the SF relation slope may steepen to $\sim2$ at higher gas surface densities \citep{Sharon2013, Rawle2014, Hodge2015, Tacconi2018}. Hence, we have adopted a `two slope' relation where star formation efficiency increases with increasing \molh\ density (labeled as \citetalias{Bigiel2008}2), and a `single slope' relation where star formation efficiency remains linearly related to \molh\ surface density (labeled as \citetalias{Bigiel2008}1; see Fig. 1 and Eqn. 6 in \citetalias{Somerville2015}). In addition, we included results from using the widely adopted `classic' cold-gas-based Kennicutt-Schmidt \citepalias{Kennicutt1998} SF recipe in our comparison \citep{Schmidt1959, Schmidt1963, Kennicutt1989, Kennicutt1998}. 

Our model uses a set of physically motivated, phenomenological and empirical recipes to track the evolution of a wide range of global physical properties of galaxies. These standard recipes include cosmological accretion and cooling, stellar-driven winds, chemical evolution, black hole growth and feedback, and galaxy mergers. A partial list of physical properties tracked in our model include stellar mass, star formation rate, masses of multiple species of gas (ionized, atomic, molecular), and stellar and gas phase metallicity. The star formation and chemical evolution histories are combined with stellar population models and a simple prescription for dust attenuation to compute predictions of observable properties (rest- and observed-frame luminosity in any desired filter). Throughout this work, we use the SSP models of \citet{Bruzual2003} with the Padova1994 \citep{Bertelli1994} isochrones and assume a universal Chabrier stellar initial mass function \citep[IMF;][]{Chabrier2003a}.

These models have been extensively tested at lower redshifts ($z \lesssim 6$) in previous works. For instance, \citetalias{Somerville2015} and \citetalias{Popping2014} present results from $z \sim 0$--6 and showed that the model predicted physical properties generally agree with observations; these properties includes star formation rate, specific star formation rate, atomic and molecular gas density, stellar and cold gas metallicity, stellar mass function, stellar-to-halo mass ratio. Free parameters in our models are calibrated to a subset of $z\sim0$ observations, including stellar-to-halo mass ratio, stellar mass function, stellar mass-metallicity relation, cold gas fraction versus stellar mass relation for disk-dominated galaxies, and the black hole mass vs. bulge mass relation. (see Appendix in \citetalias{Yung2019} for details). Without retuning these parameters to match observational constraints at higher redshifts, \citetalias{Yung2019} has shown that the predicted UV luminosity functions and the cosmic star formation rate (CSFR) agree well with observational constraints up to $z \sim 8$.

In \citetalias{Yung2019}, we found that the key model parameters that have strong effects on the predicted rest-frame UV luminosity are the SF timescale ($\tau_{*,0}$), which effectively characterizes the gas depletion time as $\Sigma_\text{SFR} \propto \tau_{*,0}^{-1}$, and stellar feedback relation slope ($\alpha_\text{rh}$), which characterizes the dependence of the mass loading factor of cold gas ejected by stellar feedback on halo circular velocity:
\begin{equation}
    \eta_{\rm out} = \dot{m}_\text{out}/\dot{m}_{*} = \epsilon_\text{SN} \left(V_0 / V_\text{c} \right)^{\alpha_\text{rh}}
\end{equation}
where $\dot{m}_\text{out}$ is the rate at which cold gas is ejected from the ISM by stellar feedback, $\dot{m}_*$ is the star formation rate, $V_\text{c}$ is the circular velocity of the galaxy, normalized by an arbitrary constant $V_0 = 200$\kms, and $\epsilon_\text{SN}$ and $\alpha_\text{rh}$ are tunable free parameters referred to as the SN feedback efficiency and SN feedback slope. In this paper we similarly explore the sensitivity of physical parameters such as stellar mass or SFR to varying these model parameters.
See \citetalias{Yung2019} and \citetalias{Somerville2015} for descriptions of model components that are not varied in this work, such as photoionization feedback, AGN feedback, disk sizes, stellar population synthesis, dust attenuation, chemical evolution, and the calibration process.

\section{Physical properties of high-redshift galaxies}
\label{sec:results}
One-point distribution functions of observable quantities or inferred physical properties are perhaps the most basic way to summarize the statistical characteristics of large populations of galaxies. These quantities have been probed independently with various tracers and observational constraints are fairly abundant. Furthermore, two dimensional distributions for these quantities reveal how one quantity scales with another. Moreover, additional clues for characterizing the physical processes and for disentangling the degeneracies among processes can be obtained via the redshift evolution of these distribution functions and scaling relations.

One of the main goals of this work is to connect observable quantities to the underlying physical properties for very high redshift galaxies. In this section, we show distribution functions for selected physical properties and scaling relations among these properties for galaxies in halos with masses ranging from $M_\text{H} \sim 10^8$--$10^{13}$ \Msun\ at $z = 4$--10. We also quantify the impact of uncertainties in our physical recipes on the resultant galaxy properties. All binned distribution functions presented in this work are available for download online at \url{https://www.simonsfoundation.org/semi-analytic-forecasts-for-jwst/}.

This section first presents a series of predicted distribution functions for key physical properties such as stellar mass ($M_*$) and star formation rate (SFR) tailored to specific types of \textit{JWST} surveys (\S\ref{sec:JWST}). We then present results of various star formation models and of systematically varying the parameters characterizing several key physical processes, and examine their impact on the predicted galaxy populations (\S\ref{sec:properties}). We also present a comprehensive comparison to predictions from other models in the literature (\S\ref{sec:models}), and show predictions for cold gas mass ($M_\text{cold}$) and molecular gas mass ($M_\text{\molh}$) (\S\ref{sec:tracers}). Stellar masses and SFR are often inferred from galaxy broadband photometry or nebular emission lines, while the cold gas content may be estimated from CO or dust continuum emission. There are significant uncertainties and inherent assumptions in estimating these physical properties from observables, and therefore it is interesting to confront these relationships with the forward modeling predictions from theoretical models.

\subsection{Distribution functions in mock \textit{JWST} surveys}
\label{sec:JWST}
Given the rather good agreement between our fiducial models and existing observations (shown in \citetalias{Yung2019}), we can use our models to forecast the physical parameters of the populations that we expect \textit{JWST} to be able to study. In \citetalias{Yung2019}, we explored the detection of high-redshift galaxies with the \textit{JWST} NIRCam broadband filters and found that the F200W filter yields the highest number of detections across $z = 4$--10. Using the observed-frame luminosity calculated for this filter, we can easily estimate which galaxies will be detectable for a survey of a given sensitivity and area. In the results presented here, we do not include noise or other instrumental effects. We plan on including these in a future work.

As illustrated in \citetalias{Yung2019}, the unprecedented sensitivity of \textit{JWST} will be able to probe objects that are several magnitudes fainter than the current detection limit of \textit{HST} deep surveys. These objects correspond to objects with stellar masses and SFRs several orders of magnitude lower than the current limits.  In this section, we perform selections on the galaxy populations predicted by our fiducial model using the observed frame IR luminosity calculated for the NIRCam F200W filter $m_\text{F200W}$. See fig. 13 in \citetalias{Yung2019}, for predicted distribution functions for observed frame IR magnitude calculated for the F200W filter and cut-offs for the detection limits and survey area. Our calculations have taken the absorption by the intervening IGM into account. We consider three distinct observing scenarios, including representative wide, deep, and lensed surveys. The assumed depths and areas for these hypothetical surveys are summarized in Table \ref{table:survey}, where the survey areas are chosen to be similar to their legacy \textit{HST} counterparts and the depths are estimated assuming the use of the F200W filter. We assumed an average 10x magnification for a lensed survey on a cluster field, and since the survey volume is inversely proportional to the magnification, we simply assume the survey area is a 10th of what we assumed for a deep survey. For each assumed survey area, adopting redshift slices with a width $dz = 1$, we also work out the critical stellar mass above which we would expect to detect less than one galaxy in teh probed volume. In reality, of course, the upper limit on the stellar mass that will be robustly probed by \textit{JWST} will be strongly affected by field-to-field variance caused by galaxy clustering and the underlying large scale structure, and this will depend weakly on the precise field geometry. We plan to make detailed predictions for cosmic variance in \textit{JWST} fields in future works, but do not address this here. 

\begin{table}
    \centering
    \caption{Summary of assumed detection limits for the NIRCam F200W filter and survey areas for representative \textit{JWST} surveys.}
    \label{table:survey}
    \begin{tabular}{ccc}
        \hline
        Survey Type & Detection Limit & Survey Area \\
        \hline
        Wide-field & 28.6 & $\sim100$ arcmin$^2$  \\
        Deep-field & 31.5 & $2\times2.2^2$ arcmin$^2$ \\
        Lensed-field & 34.0 & $\frac{1}{10} (2\times2.2^2)$ arcmin$^2$ \\
        \hline
    \end{tabular}
\end{table}

\begin{figure*}
    \includegraphics[width=2\columnwidth]{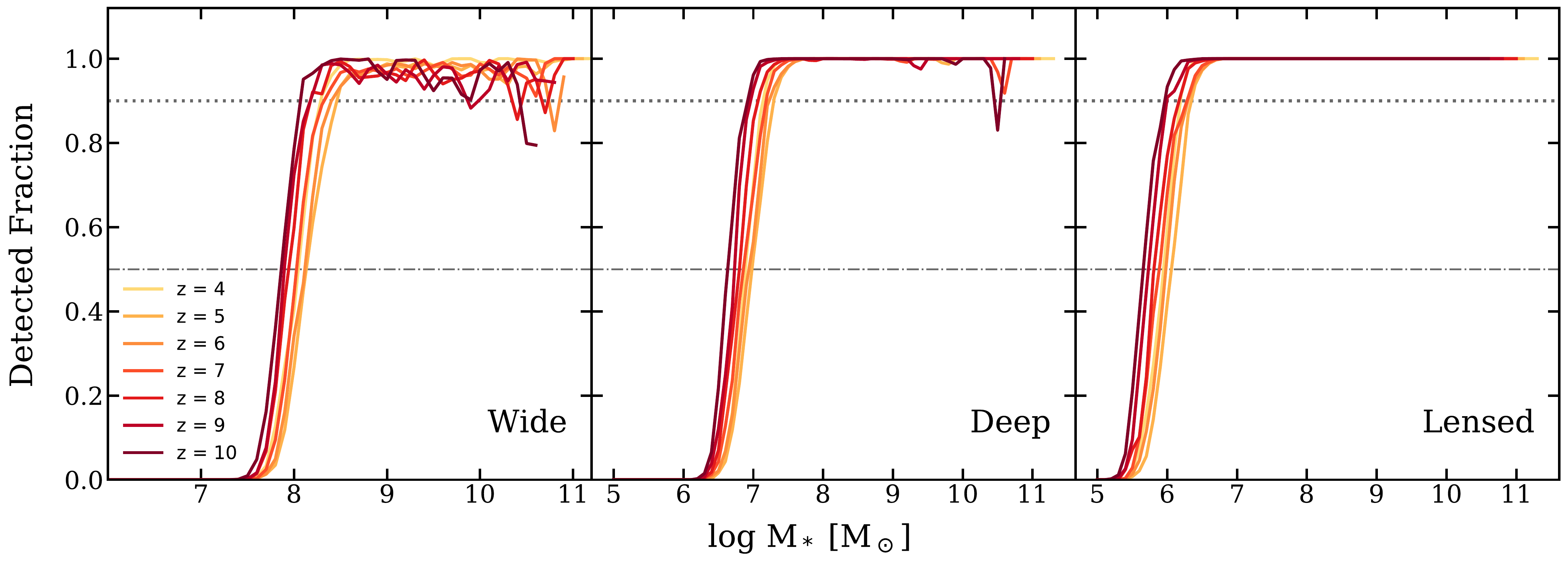}
    \caption{Fraction of galaxies detectable in wide, deep, and lensed \textit{JWST} surveys from $z=4$--10. The dotted and dot-dashed lines show detection fractions of 90\% and 50\%.}
    \label{fig:detected_fraction}
\end{figure*}

Fig. \ref{fig:detected_fraction} shows the fraction of galaxies expected to be detected in wide, deep, and lensed \textit{JWST} surveys as a function of stellar mass, using a sliding boxcar filter of width $\Delta M_* = 0.2$ in stellar mass. Stellar masses corresponding to 50 and 90\% completeness for each redshift and survey configuration can be read off from these plots. Note that the reason that the completeness does not reach a perfect value of unity even for rather massive galaxies is that some of these massive galaxies are predicted to be significantly attenuated by dust in our models, resulting in a large scatter in $m_\text{F200W}$ and $M_*$ (see fig. \ref{fig:corner_z6_trimmed}--\ref{fig:corner_z10_trimmed}).

\begin{figure}
    \includegraphics[width=\columnwidth]{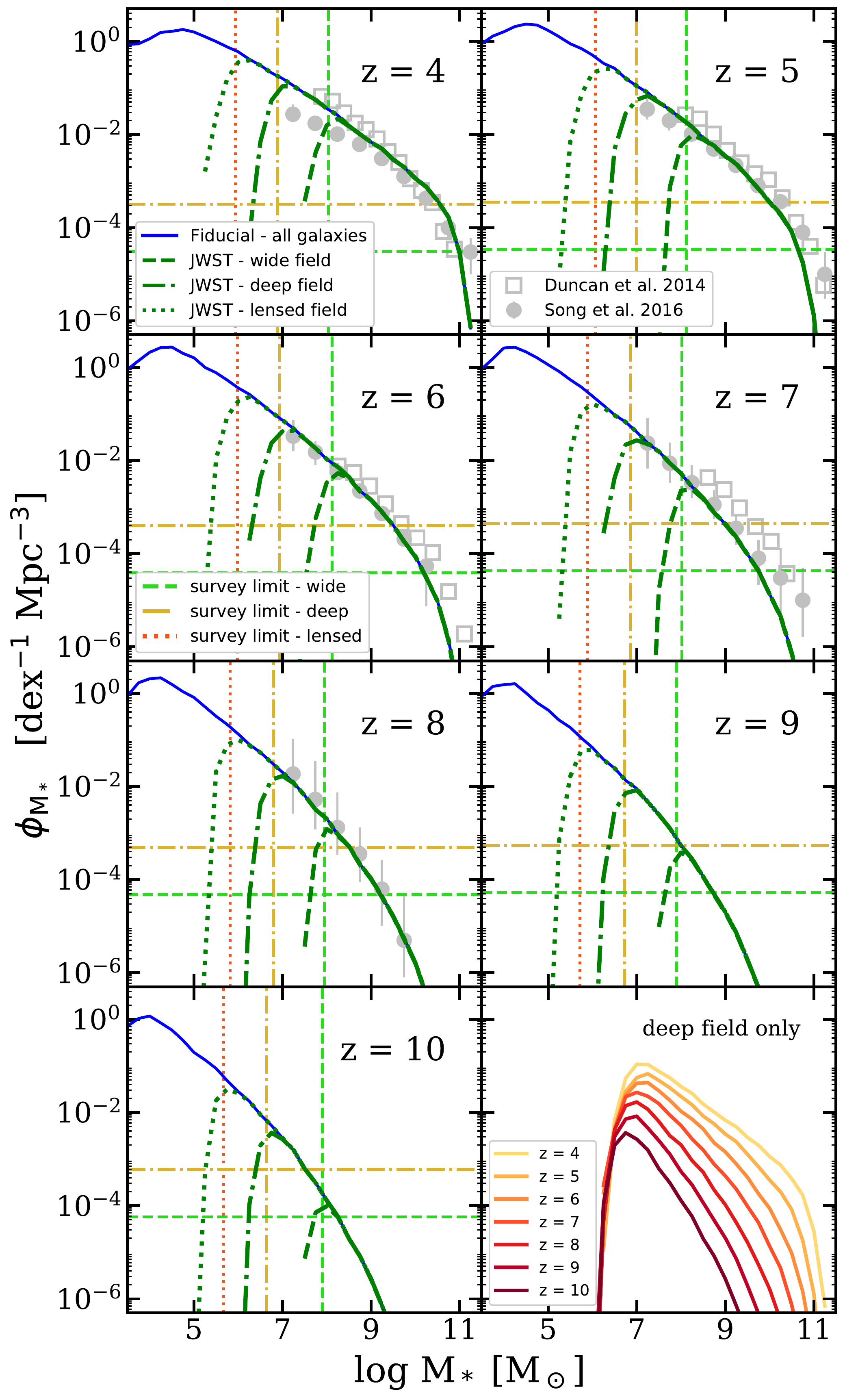}
    \caption{Predicted stellar mass functions (SMFs) and their evolution with redshift, based on our fiducial model. Results are shown for all galaxies, and for samples selected to represent wide, deep, and lensed \textit{JWST} surveys.   Observational constraints from \citet{Duncan2014} and \citet{Song2016} are shown to guide the eye. The vertical lines mark where the survey completeness reaches 50\%, and the horizontal lines mark where we expect one galaxy in the probed volume. The vertical and horizontal lines are color coded according to the survey area and depth as shown in the legend. See Table \ref{table:survey} for assumed survey specifications. }
    \label{fig:SMF_JWST}
\end{figure}

\begin{figure}
    \includegraphics[width=\columnwidth]{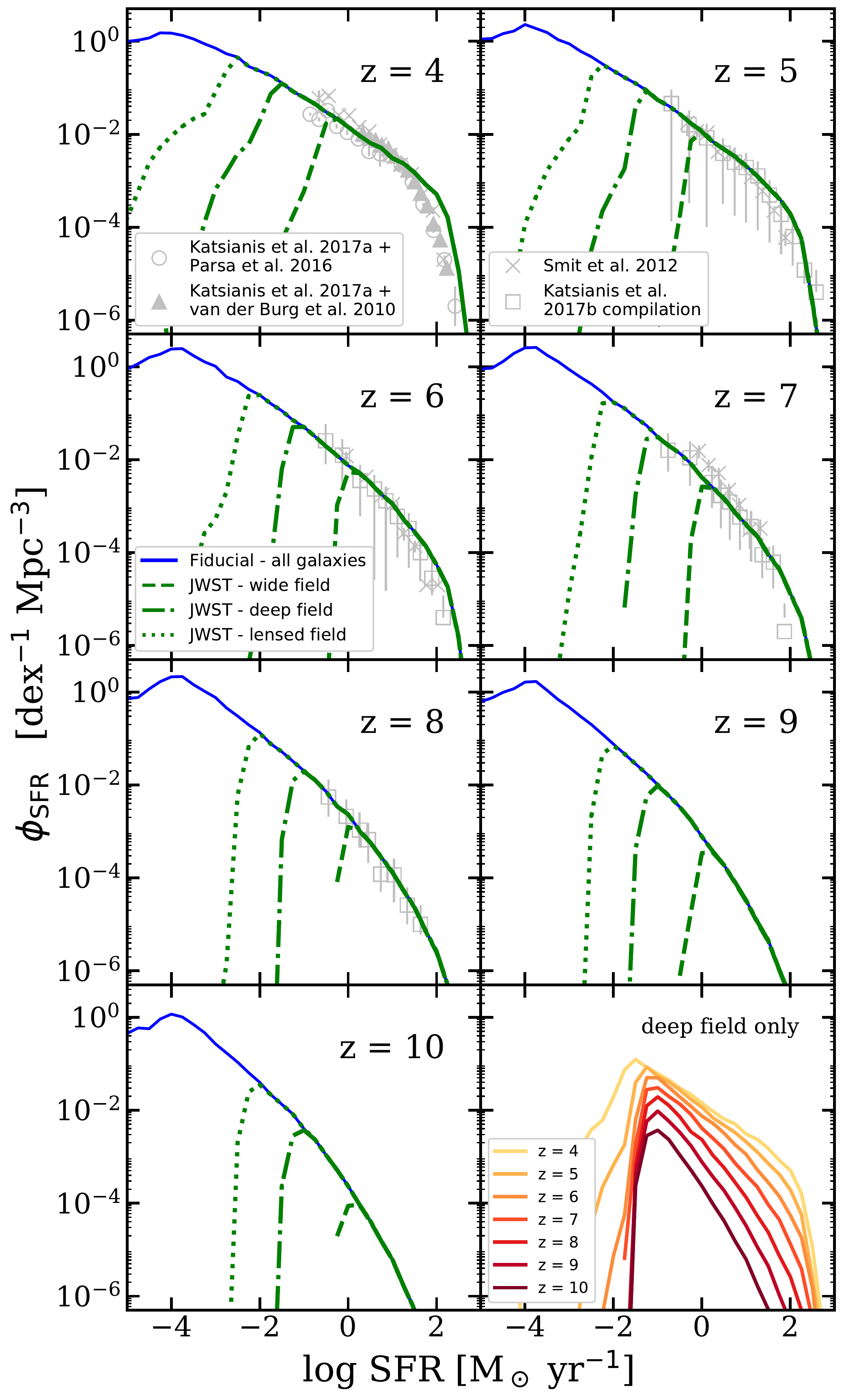}
    \caption{Predicted star formation rate functions (SFRFs) and their evolution with redshift. Results are shown for all galaxies, and for samples selected to represent wide, deep, and lensed \textit{JWST} surveys. Observational constraints are shown to guide the eye, as in Figure \ref{fig:SFRF}. See Table \ref{table:survey} for assumed survey specifications.}
    \label{fig:SFRF_JWST}
\end{figure}

Fig. \ref{fig:SMF_JWST} shows the `observable' SMFs for the three survey configurations at $z = 4$--10. We also include the full range of predictions and observational constraints from \citet{Duncan2014} and \citet{Song2016} to guide the eye. We show the stellar mass where the SMF becomes 50\% incomplete with vertical lines, and the number density where the expected number of objects in the survey drops below ten objects by horizontal lines. Thus, for a given survey area and depth, one can determine where the stellar mass function will be robustly probed by \textit{JWST} by considering the part of the function that lies above the horizontal line and to the right of the vertical line of a given color. Similarly, fig. \ref{fig:SFRF_JWST} shows the 'observable' SFRFs. We include the full range of observational constraints from \citet{Katsianis2017}, \citet{Katsianis2017a}, and \citet{Smit2012} to guide the eye. 

Although the populations that will be detected in wide-field NIRCam surveys seem to be fairly comparable to those of existing \textit{HST} deep surveys (e.g. the CANDELS Deep survey has a $H$-band detection limit of $m_\text{AB}\sim27.8$ \citep{Grogin2011}, also see Table 6 in \citetalias{Yung2019}), current stellar mass and photometric redshift estimates rely heavily on observations from the Spitzer Space Telescope. \textit{JWST} is expected to obtain much more precise photometry in the redder bands, which will also improve the photometric redshift measurements, as well as physical parameter estimates. Moreover, spectroscopic detections at $z \sim 4$--6 from NIRSpec and MIRI will provide additional constraints for these populations. It is also quite encouraging to see that lensed surveys will bring significant improvements to detecting low-stellar-mass galaxies, similar to the ultra-faint dwarf galaxies found in the local group.  Our predictions also show crudely where we can expect \textit{JWST}'s very limited field of view to cause the errorbars on the abundances of massive galaxies to become very large due to poor sampling and field-to-field variance.

In the bottom right panel of fig. \ref{fig:SMF_JWST} and \ref{fig:SFRF_JWST}, we illustrate the redshift evolution of the populations expected to be detectable in a deep-field survey. It is noteworthy that the expected stellar mass and SFR corresponding to a given completeness limit evolves rather little from $z\sim4$--10, because our models predict that high-redshift galaxies are intrinsically brighter than their low-redshift counterparts of similar masses due to their overall younger stellar populations and higher SFR. In addition to that, massive galaxies at high redshifts have higher dust-extincted luminosities due to the lower dust content. This will be briefly discussed in \S5 and further investigated in great detail in Paper III.

\subsection{Evolution of stellar mass and star formation rate distributions for galaxy populations}
\label{sec:properties}

\begin{figure}
    \includegraphics[width=\columnwidth]{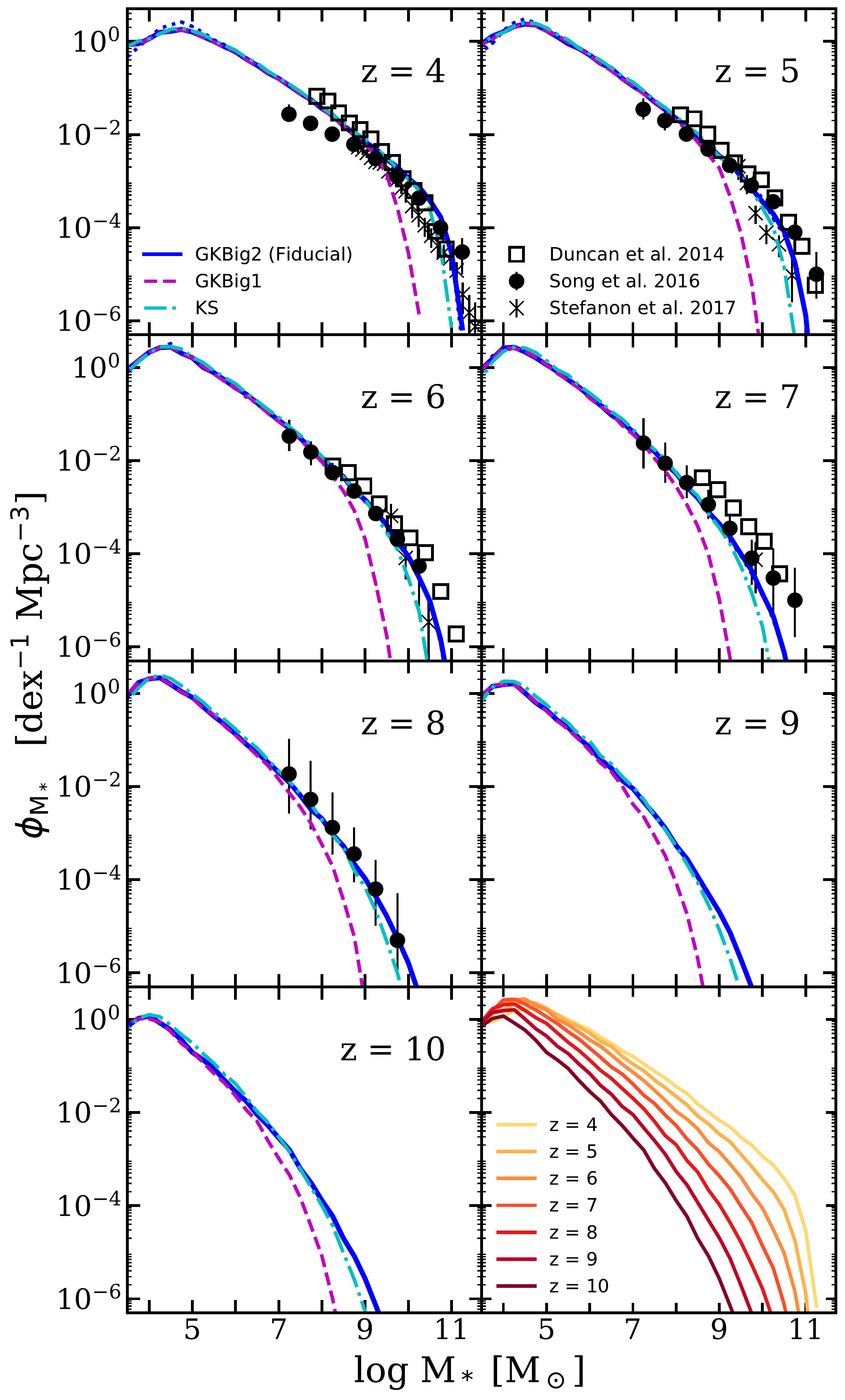}
    \caption{Predicted stellar mass functions (SMFs) and their evolution with redshift. The blue solid line shows the results of the GK-Big2 (fiducial) model, the purple dashed line shows the GK-Big1 model, and the cyan dot-dashed line shows the KS model. Our results are compared to observational constraints from \citet{Duncan2014}, \citet{Song2016}, and \citet{Stefanon2017}. We see that the GK-Big2 (fiducial) and KS model are consistent with observations, while the GK-Big1 model does not produce enough massive galaxies. We also show the case where photoionization squelching is turned off for our fiducial model with the blue dotted line. The differences are very subtle and are only visible for the very lowest mass galaxies at $z = 4$ and 5. See text for full explanation.}
    \label{fig:SMF}
\end{figure}

Fig. \ref{fig:SMF} shows the redshift evolution of stellar mass functions (SMFs) between $z = 4$--10 predicted by the three different SF models; \citetalias{Gnedin2011}-\citetalias{Bigiel2008}2 (fiducial), \citetalias{Gnedin2011}-\citetalias{Bigiel2008}1, and \citetalias{Kennicutt1998}. As shown in \citetalias{Somerville2015}, all of these models produce results that qualitatively agree with observations at $z = 0$. However, due to their differences in gas depletion time (defined as the molecular gas mass divided by the star formation rate), the predictions from these models can vary quite a lot at high redshift. As shown in Fig. 14 of \citetalias{Somerville2015}, the gas depletion time is shorter in massive galaxies at high redshift in the \citetalias{Gnedin2011}-\citetalias{Bigiel2008}2 model because of the steeper slope of the relationship between $\Sigma_\text{\molh}$ and $\Sigma_\text{SFR}$ in dense gas (see Fig. 1 and Eqn. 6 of \citetalias{Somerville2015}).

\begin{figure}
    \includegraphics[width=\columnwidth]{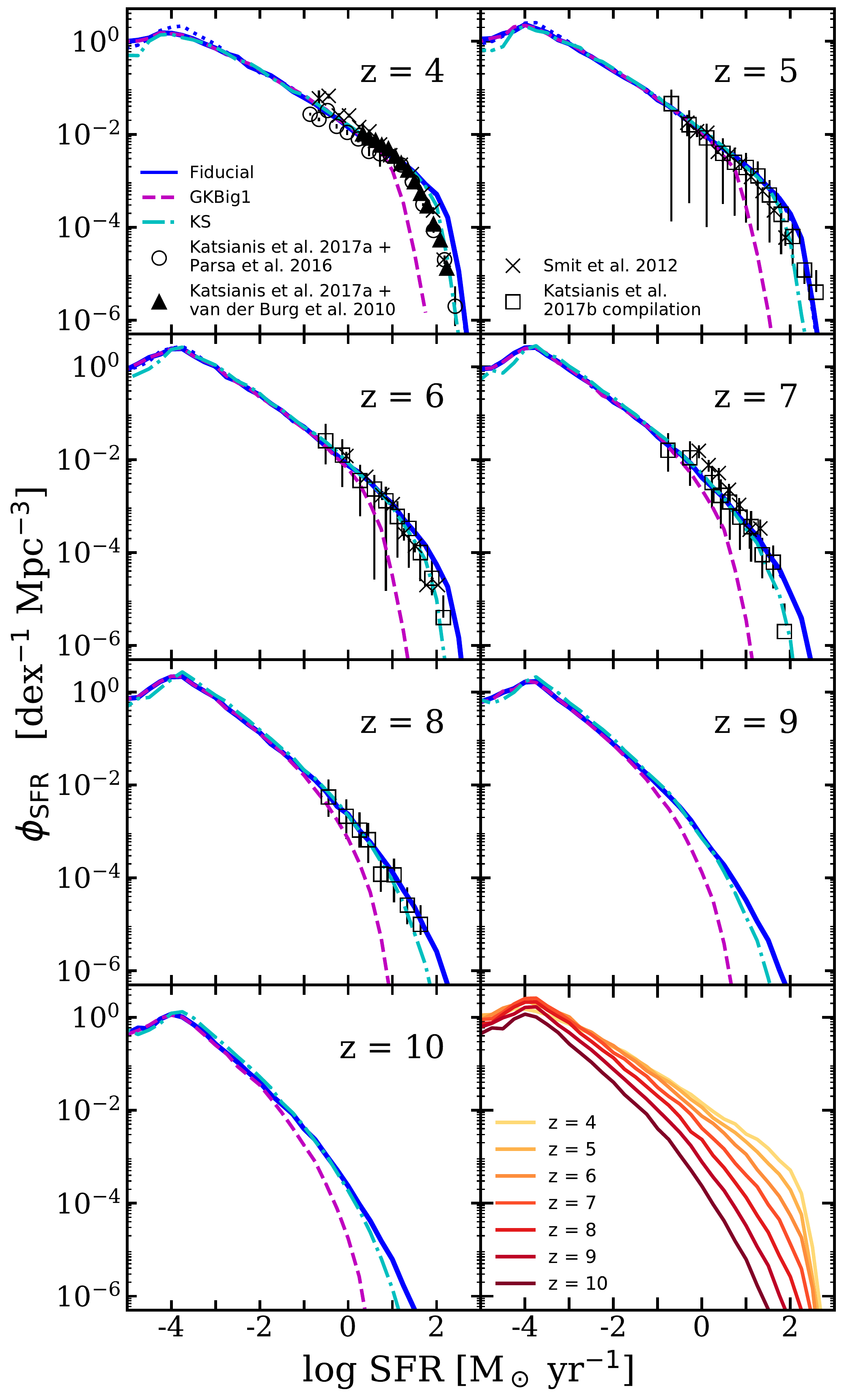}
    \caption{Predicted star formation rate functions (SFRFs) and their evolution with redshift. The blue solid line shows the results of the GK-Big2 (fiducial) model, the purple dashed line shows the GK-Big1 model, and the cyan dot-dashed line shows the KS model. Our results are compared to the \citet{Katsianis2017} study, which is based on observations from \citet{Parsa2016}  and \citet{Vanderburg2010}, and the \citet{Katsianis2017a} compilation, which is calculated based on UV LFs from \citet{Bouwens2015}, and \citet{Smit2012}. Once again, the GK-Big2 (fiducial) and KS model are consistent with observations, and the GK-Big1 model does not produce enough rapidly star forming galaxies. We also show the case where photoionization squelching is turned off for our fiducial model with the blue dotted line. The differences are very subtle and are only visible for the very lowest mass galaxies at $z = 4$ and 5. See text for full explanation.}
    \label{fig:SFRF}
\end{figure}

The low-mass end of the SMFs is quite insensitive to changing the star formation recipe, while the abundance of massive galaxies at high redshift becomes very sensitive to the scaling of the \molh\ consumption time. The reason for this is that star formation is regulated by stellar feedback in lower mass galaxies and at lower redshifts, while in massive galaxies at high redshift, the time to convert \molh\ into stars becomes the rate limiting factor in star formation. Similar behavior was seen in the response of the rest-UV LF to changes in the star formation recipe in \citetalias{Yung2019}. See \citetalias{Somerville2015} and \citetalias{Somerville2015a} for a more detailed discussion.

Our predictions are compared to observational constraints from \citet{Duncan2014} and \citet{Song2016}. It is quite encouraging that even though our model is only calibrated to $z\sim0$ observations, predictions from our fiducial model at these high redshifts still show overall very good agreement with these observational estimates.  However, at $z\sim 4$, the low-mass-end slope is significantly steeper than that reported by \citeauthor{Song2016} and seems to favor the \citeauthor{Duncan2014} observational estimates. Note that the difference between our fiducial (\citetalias{Gnedin2011}-\citetalias{Bigiel2008}2) and \citetalias{Gnedin2011}-\citetalias{Bigiel2008}1 models is the $\Sigma_\text{SFR}$-$\Sigma_\text{\molh}$ relation adopted in the SF recipe, which in the former case steepens in \molh\ dense regime. The steepening feature is crucial for reproducing the current observational constraints.

The \citet{Duncan2014} and \citet{Song2016} observational estimates of the stellar mass function show significant discrepancies with one another, particularly at the high-mass end at $z\sim 7$ and the low-mass end at $z\sim 4$. The \citeauthor{Duncan2014} study is based on the CANDELS GOODS South field, and \citeauthor{Song2016} is based on observations from the CANDELS GOODS fields, the HUDF, and other parallel fields. The main differences between their results originate from the assumed normalization and slope of the $M*$--$M_\text{UV}$ relation, which causes \citeauthor{Duncan2014} to consistently find higher stellar masses for galaxies than \citeauthor{Song2016} in faint UV bins. These two studies also treat faint objects slightly differently. \citeauthor{Duncan2014} fitted their $M*$--$M_\text{UV}$ relation to a wide stellar mass range down to $\log(M_*/M_\odot)\sim 8$, where stellar masses for galaxies with $\log(M_*/M_\odot) < 9$ are biased toward higher masses, and \citeauthor{Song2016} used a hybrid approach where high-mass galaxies are fitted individually and lower-mass galaxies are stacked. Moreover, the \citeauthor{Song2016} error bars only include random uncertainties, while the \citeauthor{Duncan2014} ones include Poison errors and photometric redshift uncertainties.

Similarly, fig. \ref{fig:SFRF} shows distribution functions for star formation rate (SFRFs) and their evolution across the same redshift range. Our results are compared to observational constraints by \citet{Katsianis2017} at $z = 4$, \citet{Smit2012} at $z = 4$--7, and \citet{Katsianis2017a} at $z = 5$--8. These constraints on SFR are determined based on a conversion between UV luminosity and SFR \citep[e.g.][]{Kennicutt1998, Smit2012}. The \citet{Katsianis2017} study is based on rest-frame UV luminosity functions from \citet{Parsa2016} and \citet{Vanderburg2010}, both of which are based on a compilation of ground- and space-based deep-field observations. The \citet{Katsianis2017a} results are calculated based on UV LFs presented in \citet{Bouwens2015}. The observational constraints shown have accounted for dust attenuation. Once again, the results from our fiducial model show good agreement with these observational constraints, and are clearly favored over the model with a constant \molh\ depletion time. Note that observational estimates of SFR derived from UV luminosity probe a timescale of approximately 100 Myr \citep{Kennicutt2012}. The SFR predictions from the SAM have been averaged over 100 Myr for an appropriate comparison.

As shown in fig. \ref{fig:HMF_check_2} in Appendix \ref{appendix:d}, the abundance of halos rises continuously towards lower halo mass. Our simulation routinely samples a wide range of halo masses down to the atomic cooling limit and halo assembly histories are traced down to 100th of the root mass. The `flattening' or `turnover' seen at the low-mass / low-SFR end of the distribution functions in our models is a physical prediction resulting from inefficient cooling, rather than from insufficient resolution as in most numerical simulations. However, we note that our model does not include \molh\ cooling or metal cooling below $10^4$ K. We find that the critical stellar mass where the turnover occurs evolves mildly over redshift, from $\log(\text{M}_*/\text{\Msun}) \sim 4$ at $z\sim 10$ to $\log(\text{M}_*/\text{\Msun}) \sim 5$ at $z\sim 4$, while the critical SFR remains nearly constant at $\log(\text{SFR}/(\text{\Msun\ yr}^{-1})) \sim -4$ over this period.  A cautionary note is that these turnovers are not representing the same galaxies. Instead, each redshift is an independent snapshot that portrays the demographics of the current galaxy population. It is also interesting to note that the SFR function is closer to a pure power-law at high redshift ($z\gtrsim 6$), and begins to develop an exponential cut-off at high SFR at $z\sim 5$.

For both SMF and SFRF predictions, we have experimented with turning off the feedback from photoionization squelching in our fiducial model, shown as a dotted line in fig. \ref{fig:SMF} and \ref{fig:SFRF}. These results show that, at least with the simple implementation adopted in our model, the effect of squelching is negligible across galaxies of all masses, even at $z = 4$. This is because the characteristic mass from the \citet{Okamoto2008} simulations is getting very close to the atomic cooling limit. Thus only the very low-mass halos are affected by squelching. Additionally, we have tested the effect of switching off AGN feedback in our models, and find that it has no discernible effect on the predictions in this redshift range (see the discussion in \citetalias{Yung2019}).

\begin{figure}
    \includegraphics[width=\columnwidth]{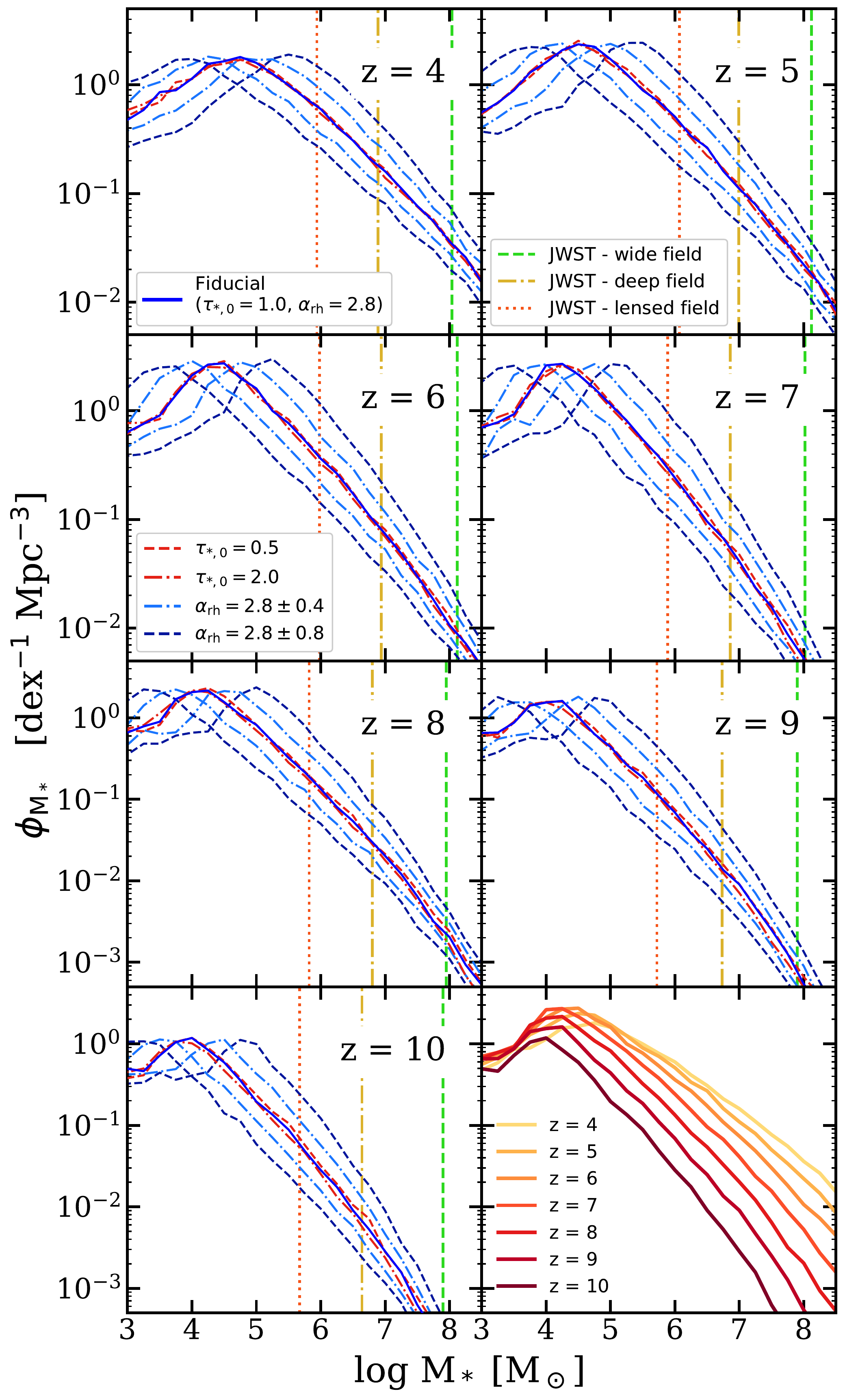}
    \caption{The low-mass end of the predicted stellar mass functions (SMFs) and their evolution with redshift. The blue solid line shows the results from our fiducial model. Red lines represent the cases where we alter the SF timescale $\tau_{*,0}$, where dashed and dot-dashed lines are $\tau_{*,0} = 0.5$ and $\tau_{*,0} = 2.0$, respectively. The light-blue lines show cases where the mass-loading factor of stellar driven winds $\alpha_\text{rh}$ is altered, where $\alpha_\text{rh} = 2.8\pm0.4$ (dot-dashed) and $\pm0.8$ (dashed). Blue dot-dashed lines show the cases where we let $\alpha_\text{rh} = 2.4$ (above) and 3.2 (below), and light blue dashed lines show the cases where we let $\alpha_\text{rh} = 2.0$ (above) and 3.6 (below). Red lines represent the cases where we increase or decrease $\tau_{*,0}$ by a factor of 2 from its fiducial value of unity; dashed and dot-dashed lines are $\tau_{*,0}$ = 0.5 and $\tau_{*,0}$ = 2.0, respectively. The vertical dashed lines represent survey completeness of 50\% for example \textit{JWST} surveys similar to legacy\textit{HST}counterparts; see Table \ref{table:survey} for details. The last panel summarizes the evolution of the low-mass end of the SMFs predicted by the fiducial model. We see that both the slope of the low-mass SMF and the location of the turnover are strongly affected by the model for stellar driven winds, but are not significantly affected by variations in the SF timescale.}
    \label{fig:SMF_variable_faint}
\end{figure}

\begin{figure}
    \includegraphics[width=\columnwidth]{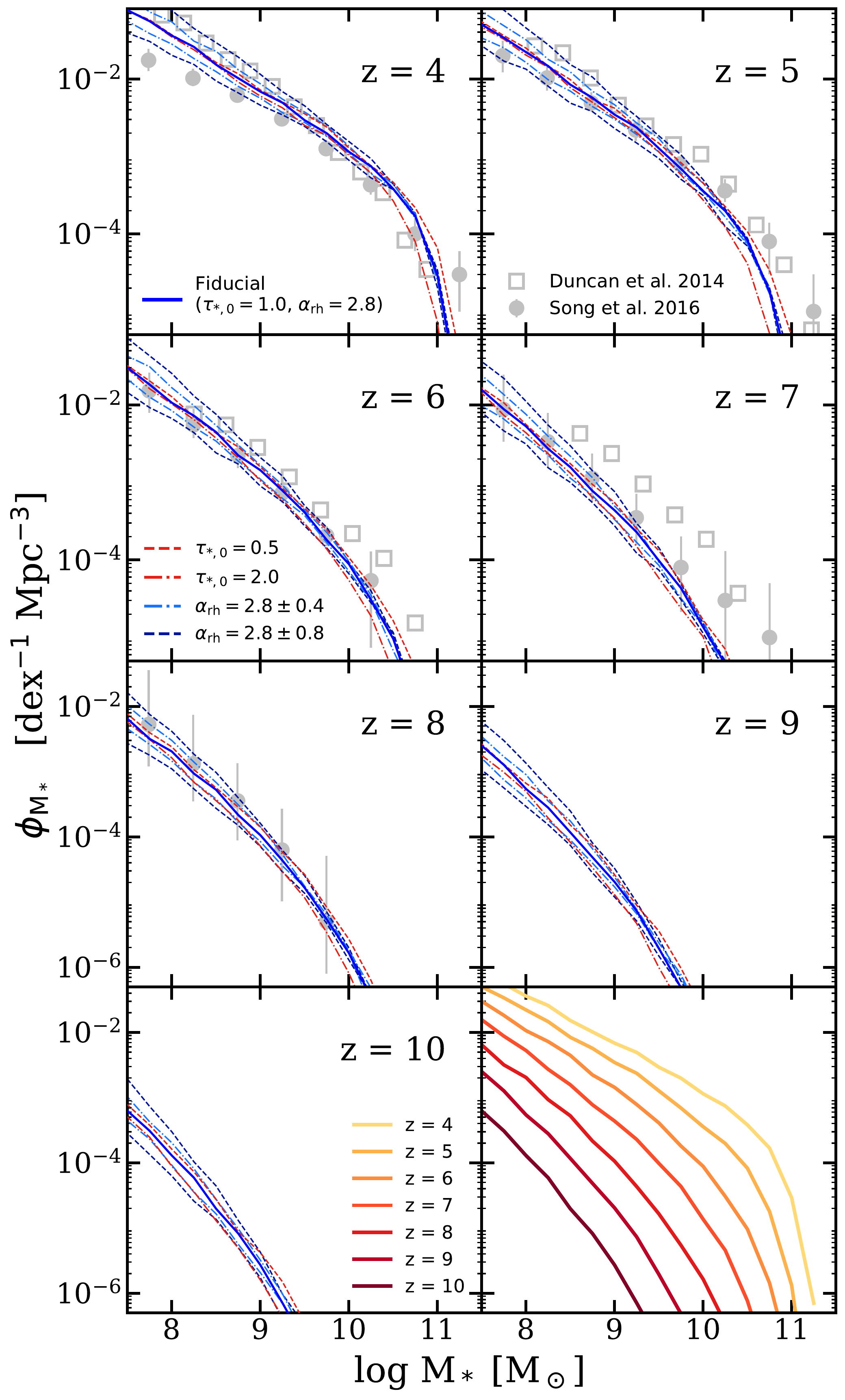}
    \caption{The massive end of the predicted stellar mass functions (SMFs) and their evolution with redshift. See caption of fig. \ref{fig:SMF_variable_faint} for legend details. The gray circle and open square symbols show observational constraints from \citet{Duncan2014} and \citet{Song2016}, respectively. The last panel summarizes the evolution of the massive end of the SMFs predicted by the fiducial model. Varying the SF timescale has the greatest impact on very massive galaxies at high redshift, while varying the stellar wind model has a smaller effect in this regime.}
    \label{fig:SMF_variable_bright}
\end{figure}

In a similar spirit as the experiments done in \citetalias{Yung2019}, we systematically varied the stellar feedback slope $\alpha_\text{rh}$ and the SF timescale $\tau_{*,0}$ within a sensible range in an attempt to quantify the effects of these uncertainties in the key model parameters.  Here, we show low-mass and massive galaxy populations separately in fig. \ref{fig:SMF_variable_faint} and fig. \ref{fig:SMF_variable_bright}, respectively.

We showed in \citetalias{Yung2019} that the low rest-frame UV luminosity galaxy populations are very sensitive to the efficiency of stellar driven winds. Here, we show a similar effect for stellar mass, where we explore four values of $\alpha_\text{rh} = 2.0$, 2.4, 3.2, and 3.6, where larger values of $\alpha_\text{rh}$ imply a steeper dependence of mass loading factor on galaxy circular velocity, resulting in more gas being ejected from galaxies in low mass halos (see \S2.3 in \citetalias{Yung2019} for details). Note that these alternative values will not necessarily reproduce observations at $z \sim 0$; we are adopting these values as an attempt to explore alternative scenarios where the strength of feedback may have been stronger or weaker at early times. In our model, feedback from stellar winds is modeled using a physically motivated, redshift-independent recipe. This feedback mechanism could be sensitive to many intertwined physical properties of galaxies, which could lead indirectly to an effective evolution with cosmic time.  The feedback efficiency is poorly constrained especially in low-mass galaxies and at high redshifts due to the challenges of obtaining direct observational constraints. Henceforth, without precluding the possibility of having an effectively evolving $\alpha_\text{rh}$, we have experimented with adopting different constant values of $\alpha_\text{rh} = 2.8 \pm 0.4$ and $\pm 0.8$. On the other hand, $\tau_{*,0}$ effectively acts as a normalizing factor for the SF relation, which converts the surface density of molecular hydrogen $\Sigma_\text{\molh}$ to the surface density of SFR $\Sigma_\text{SFR}$ (see \S2.1 in \citetalias{Yung2019} for details). Larger $\tau_{*,0}$ represents a longer SF timescale or, effectively, a longer gas depletion timescale, yielding a lower star formation rate per unit surface \molh\ density, and vice versa. To explore the effect of uncertainties in the observationally-motivated, empirical SF recipe, we experiment with increasing and decreasing $\tau_{*,0}$ by a factor of two from its fiducial value of unity (which corresponds to the observed normalization).

In fig. \ref{fig:SMF_variable_faint}, we see that the low-mass-end slope $\alpha_\text{rh}$ becomes steeper when $\alpha_\text{rh}$ is decreased to 2.4 and 2.0, due to the less efficient mass ejection by stellar feedback. The slope flattens when $\alpha_\text{rh}$ is increased to 3.2 and 3.6. In the least massive populations, we see a shift in the `turn over' mass when we vary the feedback efficiency. This is because the turnover in our models always occurs roughly at a fixed halo circular velocity (temperature) corresponding to the atomic cooling limit, but changing the feedback prescription changes the relationship between halo mass or velocity and galaxy stellar mass. We also find that the low-mass populations are insensitive to the adopted value of $\tau_{*,0}$ (see \citetalias{Somerville2015} for a detailed discussion of the physical reasons for this).

Similarly, in fig. \ref{fig:SMF_variable_bright}, we explore the impact of adopting a range of values for $\alpha_\text{rh}$ and $\tau_{*,0}$ on the massive galaxy populations. In this regime, the results are more degenerate, showing significant dependence on both parameters, although SF timescale has a stronger effect on more massive galaxies and the SN feedback has a greater effect on low-mass objects.
We note that tuning $\tau_\text{*,0}$ alone to alter the bright end prediction will result in a change in predicted gas fraction. Thus observational constraints on gas content in galaxies at high redshift can help to break these degeneracies.

As noted before, the current observational constraints on stellar masses of faint objects at high redshift are highly uncertain, as reflected in the discrepancies in current estimates in the literature. Our models can easily accommodate the range of scenarios presented by current observations with variations in parameters that are well within the observational and theoretical uncertainties. Our results illustrate how future measurements with \textit{JWST} and wide field surveys with LSST and WFIRST will be complementary in constraining different physical processes in galaxy formation at extreme redshifts.

\subsection{Comparison with other models}
\label{sec:models}
In this subsection, we compare the SMFs and SFRFs predicted by our fiducial model to a representative collection of theoretical predictions, including empirical models, semi-analytic models, and cosmological hydrodynamic simulations from the literature.  First, we very briefly summarize the specifications of these models and simulations. We note that these models adopt different cosmological parameters, very different approaches for modeling the baryonic physics, and different approaches for calibration. We compare the results at face value without attempting to correct for any of these differences. Furthermore, providing the full details about these models or attempting to understand the sources of differences in their predictions is beyond the scope of this work. We refer the reader to the included references for each of these simulations for full details.

A number of numerical simulations are included in our comparison; \textsc{BlueTides} is a large-volume cosmological hydrodynamic simulation that focuses on the high-redshift universe, with a box that is 400 Mpc $h^{-1}$ on a side, resolving galaxies with $M_* \gtrsim 10^8$\Msun\ toward the end of their simulation, which stops at $z \sim 8$. They presented predictions for both SMFs and SFRFs \citep{Wilkins2017}. The Illustris simulation has a box 106.5 Mpc on a side with dark matter particle of mass $6.3\times10^6$\Msun\ \citep{Genel2014}.  The Evolution and Assembly of Galaxies and ther Environment (\textsc{Eagle}) simulations is a suite of cosmological hydrodynamical simulations of simulated volumes ranging from 25 to 100 cMpc at various mass resolution \citep{Schaye2015}. In our comparison, we used the stellar mass functions from their Ref-L100N1504 and Recal-L025N0752 runs, which have box sizes of 100 cMpc and 25 cMpc on a side and dark matter particle mass of $1.21\times10^6$ \Msun\ and $9.70\times10^6$ \Msun, respectively \citep{Furlong2015}.  The Feedback in Realistic Environments (FIRE) simulations are a suite of `zoom-in' simulations extracted from cosmological volume simulations, for which regions are `resimulated' at higher resolution with detailed physical processes incorporated \citep{Hopkins2014}. We use SMFs from the FIRE-2 simulations \citep{Hopkins2018, Ma2017}, which are obtained by weighting the results from the zoom-ins with cosmological halo mass functions. Due to space limitations, other simulations, such as the Renaissance Simulations \citep{OShea2015}, the FirstLight project \citep*{Ceverino2017, Ceverino2018, Ceverino2019}, and SPHYNX \citep*{Cabezon2017}, as well as semi-analytic model results from \citet{Dayal2014}, are omitted from our comparison.

We also compare with predictions from the Dark-ages Reionization and Galaxy formation Observables from Numerical Simulations (DRAGONS) project, which consists of a SAM \textsc{meraxes} \citep{Mutch2016} that is built on top of the \textit{Tiamat} suite of $N$-body simulations of dark matter particle of mass $\sim 2.64 \times 10^6$ $ h^{-1}$ Mpc \citep{Poole2016}. We show the SMFs presented in \citet{Qin2017}.

\begin{figure}
    \includegraphics[width=\columnwidth]{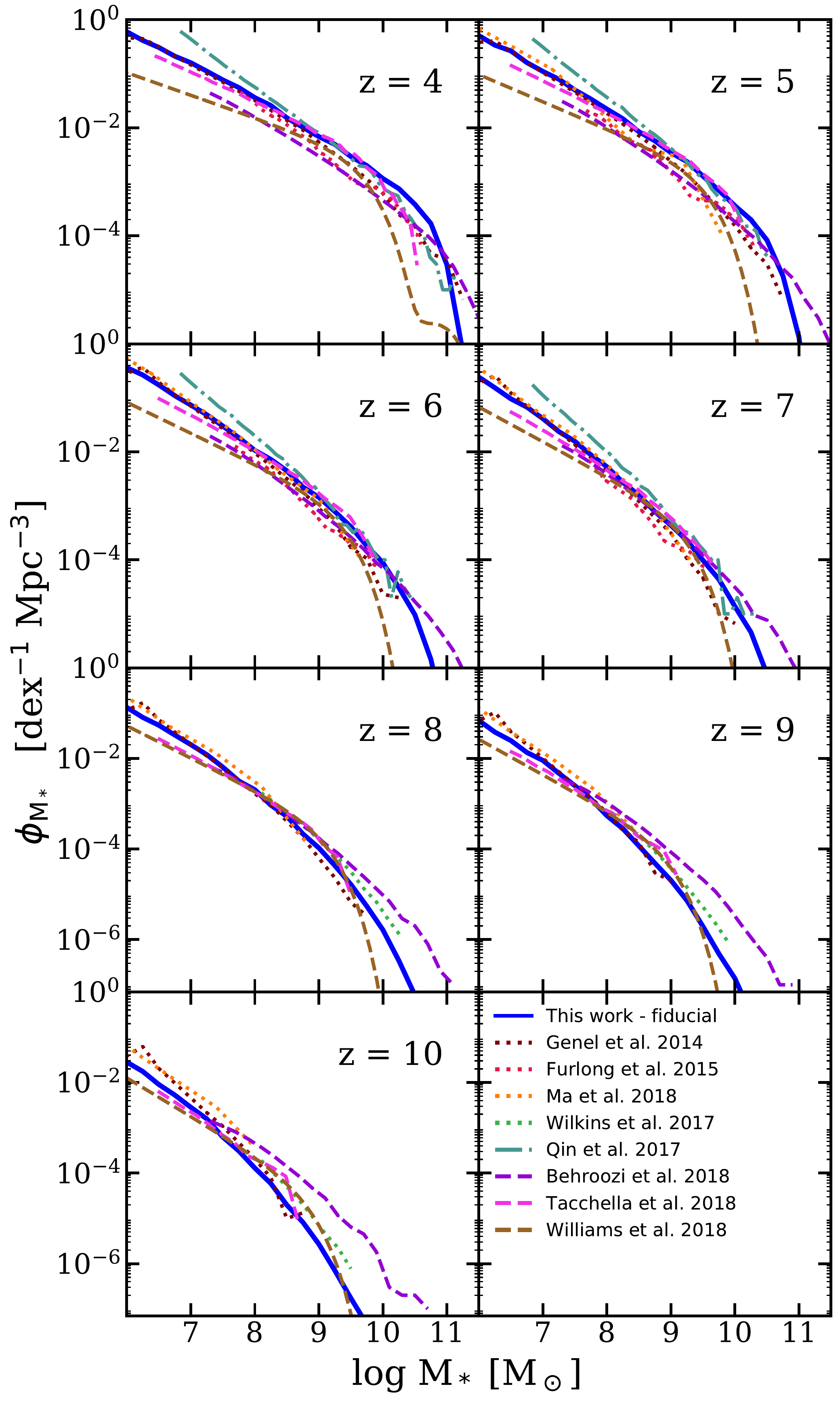}
    \caption{Redshift evolution of the SMF between $z = 4$--10 predicted by our fiducial model, compared to a compilation of other theoretical studies, including empirical models \textsc{UniverseMachine} \citep{Behroozi2018}, \citet{Tacchella2018}, and JAGUAR \citep{Williams2018}; numerical simulations from Illustris \citep{Genel2014}, \textsc{Eagle} \citep{Furlong2015}, and \textsc{BlueTides} \citep{Wilkins2017}, and semi-analytic models from DRAGONS \citep{Qin2017}. See text for details. Overall, the agreement between different methods is fairly good.}
    \label{fig:smf_models}
\end{figure}

\begin{figure}
    \includegraphics[width=\columnwidth]{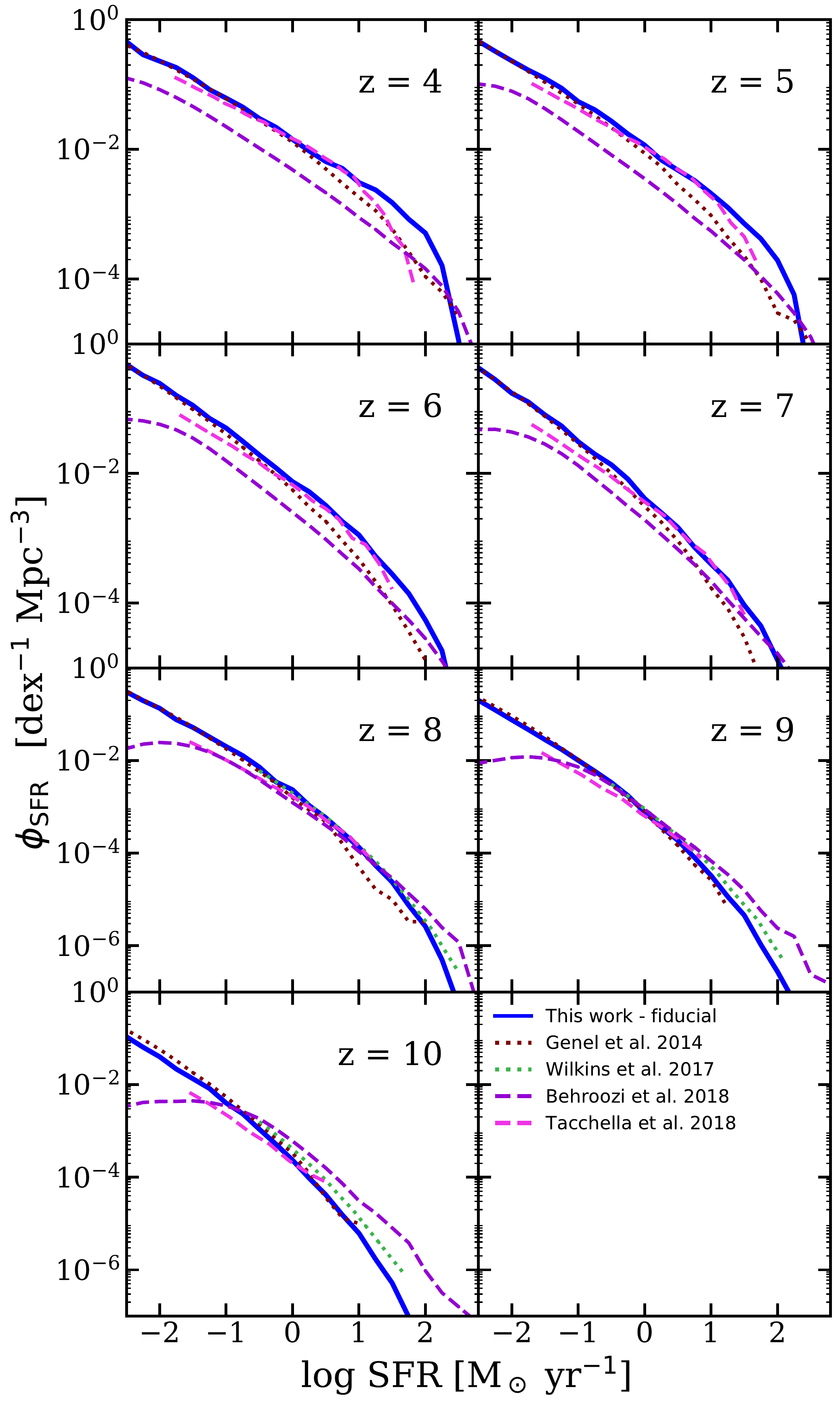}
    \caption{Redshift evolution of the SFRF between $z = 4 - 10$ predicted by our fiducial model, compared to a compilation of other theoretical studies, including empirical models \textsc{UniverseMachine} \citep{Behroozi2018} and \citet{Tacchella2018}, numerical simulations from Illustris \citep{Genel2014}, \textsc{Eagle} \citep{Furlong2015}, \textsc{BlueTides} \citep{Wilkins2017}. See text for details.}
    \label{fig:sfrf_models}
\end{figure}

We include comparisons with several sub-halo abundance matching (SHAM), empirical, and semi-empirical models.
The \textsc{UniverseMachine} \citep{Behroozi2018} obtained dark matter halo populations and properties from the Bolshoi-Planck dark matter simulations \citep{Klypin2016, Rodriguez-Puebla2016}, which simulate a volume of 250 Mpc $h^{-1}$ on a side containing $2048^3$ dark matter particles of $1.5 \times 10^8$\Msun. Within these dark matter halos, galaxy populations are constructed to fit the SFR--$v_\text{Mpeak}$ relation, quenching--$v_\text{Mpeak}$ relation, quenching/assembly history correlation in the local universe ($z\sim 0$), and dust content at high redshifts ($z\sim4$--10) from a wide range of observational constraints. In this work we use the \textsc{UniverseMachine} Early Data Release (EDR) catalogs\footnote{\url{http://behroozi.users.hpc.arizona.edu/UniverseMachine/EDR/umachine-edr.tar.gz}}. The \citet{Tacchella2018} model uses dark halo merger trees extracted from the \textit{Copernicus complexio Low Resolution} (\textsc{color}) simulations \citep{Hellwing2016, Sawala2016}, which contain $1620^3$ dark matter particles of $6.196 \times 10^6$ \Msun$h^{-1}$ in a periodic volume of 70.4 Mpc $h^{-1}$ on a side. The only observational constraint this model is calibrated to is the UV LF observed at $z = 4$, where the star-formation efficiency is treated as a free parameter. Their results are then checked against the measured cosmic star formation rate, $M_\text{UV}$--$M_*$ relation, stellar mass function, etc. The JADES Extragalactic Ultra-deep Artificial Realization \citep[JAGUAR;][]{Williams2018} model provides mock catalogs specifically for upcoming \textit{JWST} observations. JAGUAR makes no assumptions about the underlying halo populations. Instead, it directly maps observable properties, such as $M_\text{UV}$, to a number of physical properties using simple relations from observations, such as the $M_\text{UV}$--$M_*$ relation, $\beta_\text{UV}$--$M_\text{UV}$ relation, and spectroscopic properties using SED fitting to \textit{3D-HST} catalogs \citep{Skelton2014, Momcheva2016}. The semi-empirical model \textsc{Emerge} \citep{Moster2018} is very similar to the \textsc{UniverseMachine} models but is not included in our comparison, again for reasons of space limitations.

In fig. \ref{fig:smf_models}, we compare our predicted SMFs to the predictions of Illustris, \textsc{Eagle}, DRAGONS,  \textsc{BlueTides}, \textsc{UniverseMachine}, \citeauthor{Tacchella2018}, and \citeauthor{Williams2018}. We note that some of these simulated SMFs are based on the \citet{Salpeter1955} IMF and the \citet{Kroupa2001} IMF, and are converted to Chabrier IMF by adding -0.21 and -0.03 dex, respectively \citep{Madau2014}. In general, most of these predictions are within 0.5 dex from our predictions within the mass range where objects are well-resolved (in the low-mass end) and well-sampled (in the massive end) in the given simulation. We also note that the excellent agreement for $M_* \gtrsim 9$ at $z \sim 4$--7 may be to some extent by construction, since all simulations are matching the available observational constraints either actively (by calibration) or passively (used as crosschecks). Our models agree quite well overall with the predictions from numerical hydrodynamic simulations, including \textsc{BlueTides}, Illustris, and FIRE, over a broad stellar mass and redshift range. It is intriguing that these models, which produce converging predictions on the massive end, are producing rather different predictions for the low-mass populations. For instance, the most optimistic model, DRAGONS, and the least optimistic model, \citeauthor{Williams2018} predict the abundance of $M_* = 10^7$\Msun\ galaxies differs by $>1$ dex at $z=7$, and the difference becomes even larger at low redshifts. The comparison also shows that there is a shortfall of massive objects in the JAGUAR predictions. Note that in this redshift range, the \citeauthor{Williams2018} model does not explicitly match their SMFs to observations, but rather forward modeled them to match UV LFs adopting simple assumptions for the $M_*$--$M_\text{UV}$ and $\beta_\text{UV}$--$M_\text{UV}$ relations based on local observations.

The SMFs at high redshift predicted by the \textsc{UniverseMachine} tend to have a higher normalization than previous results due to improved treatment of star formation history priors, which require more stellar mass especially at $z \sim 8$ (P. Behroozi, priv. comm.). On the other hand, the lower mass limit in halo mass in the \textsc{UniverseMachine} is $\sim10^{10.5}$--$10^{10.7}$\Msun\, below which merger trees are not sufficiently resolved to result in accurate galaxy properties. We note that both the \citeauthor{Tacchella2018} model and the \textsc{UniverseMachine} are not explicitly calibrated to match observed SMF constraints at $z > 4$. Due to the specific set of assumptions and choices made in their models, though both of these models are able to reproduce observations in the low-redshift universe, their predictions in the poorly constrained high-redshift differ significantly. Investigating and understanding these differences in detail is beyond the scope of this work, and we refer the reader to the relevant papers for full details.

Similarly, in fig. \ref{fig:sfrf_models}, we compare our predicted SFRFs to Illustris, \textsc{BlueTides}, \textsc{UniverseMachine}, and \citeauthor{Tacchella2018}. We note that some of the SFRs predicted by these simulations are based on the \citet{Salpeter1955} IMF and the \citet{Kroupa2001} IMF, and are converted to Chabrier IMF by adding -0.20 and -0.03 dex, respectively, for this comparison \citep{Madau2014}. At first glance, predictions by Illustris and \citeauthor{Tacchella2018} seem to agree with predictions from our fiducial model quite well for the abundance of galaxies with $\text{SFR} \sim 0.1$ \Msun\ yr$^{-1}$ across redshifts. However, they diverge for $\text{SFR} \gtrsim 1$ \Msun\ yr$^{-1}$. The SFR predicted by \textsc{UniverseMachine} at $z \lesssim 7$ are consistently lower than other predictions by $\sim0.5$ dex and are higher at $z \gtrsim 9$ for similar reasons to those noted above.

In summary, it is encouraging that physically motivated, \textit{a priori} models make similar predictions for these fundamental quantities at redshifts where they were not calibrated to observations, in spite of containing very different modeling approaches to baryonic processes. The differences between these models and empirical and semi-empirical models may yield interesting insights into the relationship between galaxy observable properties and halo properties. We pursue this further in a later section of this paper.

\subsection{Cold gas content of high redshift galaxies}
\label{sec:tracers}
We learned that both cold gas and molecular gas are important tracers for star formation activity from observing nearby galaxies \citep[e.g.][]{Kennicutt1998, Bigiel2008}. Empirical relations based on these observations are then extensively incorporated in galaxy formation models, including both conventional numerical simulations and semi-analytic models. Therefore, in addition to $M_*$ and SFR, the distributions of cold gas mass ($M_\text{cold}$) and molecular gas mass ($M_\text{\molh}$) and their evolution can also provide important constraints for disambiguating stellar feedback processes. Historically, it has been somewhat challenging for models to simultaneously match the observed gas fraction, stellar fraction, and stellar metallicity (see \citetalias{Somerville2015a} for a detailed discussion).

The gas depletion timescale is the timescale for converting all the mass in the cold gas reservoir into stars. The three SF recipes adopted in our SAM can be effectively understood as different redshift evolution scenarios for gas depletion time. For instance, the GK-Big1 model corresponds to a constant \molh\ depletion time and the GK-Big2 model predicts a gas depletion time that changes with time as the galactic discs gradually become less dense over time.  The evolution of these quantities predicted by our models at $z \sim 0$--6 have been studied in depth in \citetalias{Somerville2015} and \citetalias{Popping2014}, which found that these models produce results that are in qualitative agreement with observed evolution in galaxies at $z\lesssim 6$. In this work, we extend the investigation to $z = 10$. Though constraints on the gas content of galaxies at such extreme redshifts are not currently available, future observational programs and observational facilities may be able to obtain them. Moreover, examining the changes in the predictions for gas content provides complementary insights into our model results.

\begin{figure}
    \includegraphics[width=\columnwidth]{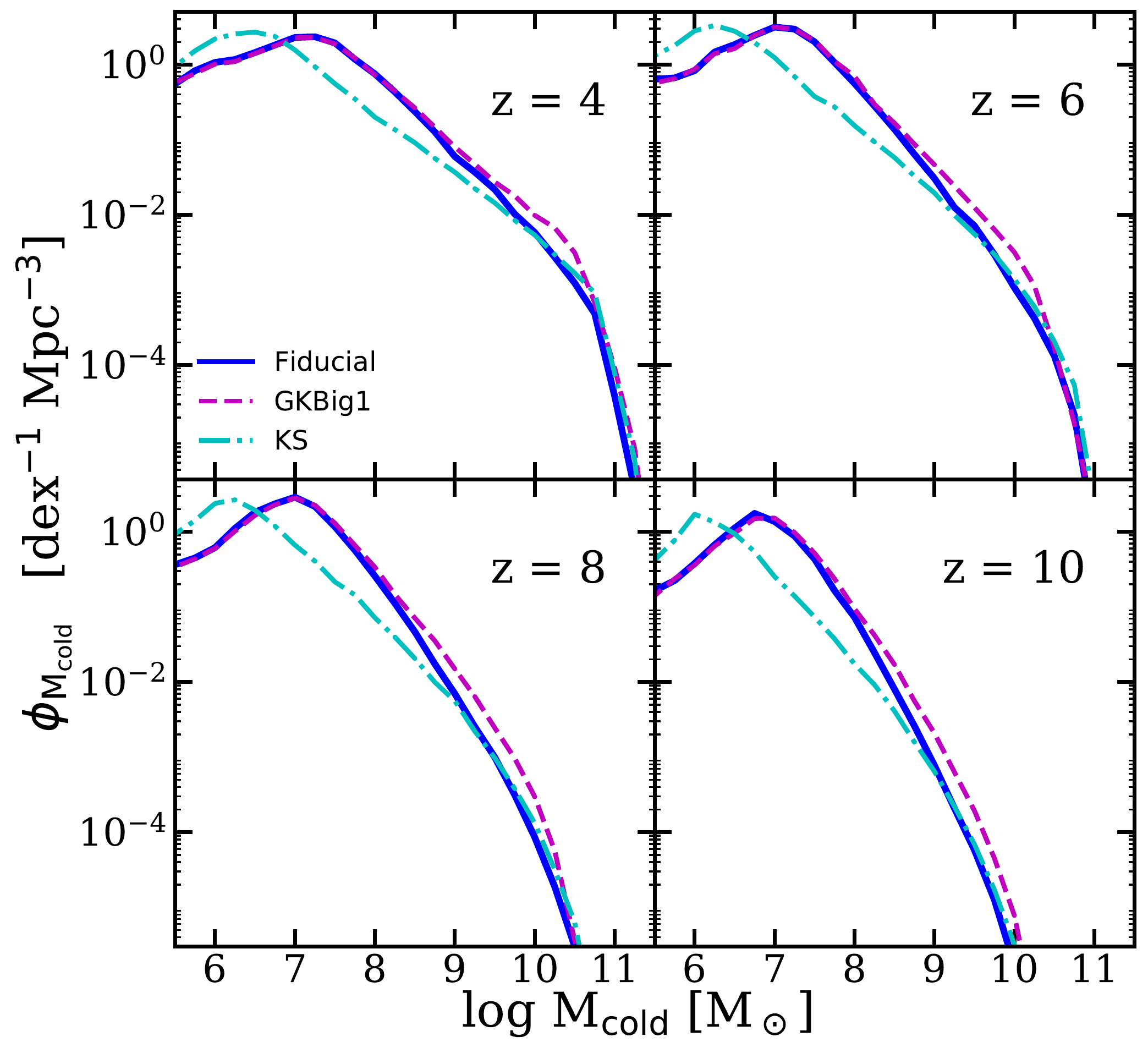}
    \caption{Predicted distribution functions for cold gas $M_\text{cold}$ at $z = 4$, 6, 8, and 10. The blue solid line shows the results of the GK-Big2 (fiducial) model, the purple dashed line shows the GK-Big1 model, and the cyan dot-dashed line shows the KS model. While these models all made very similar predictions for the low-mass end of the SMF, they make rather different predictions for the cold gas mass function at low gas masses. }
    \label{fig:CGMF}
\end{figure}

\begin{figure}
    \includegraphics[width=\columnwidth]{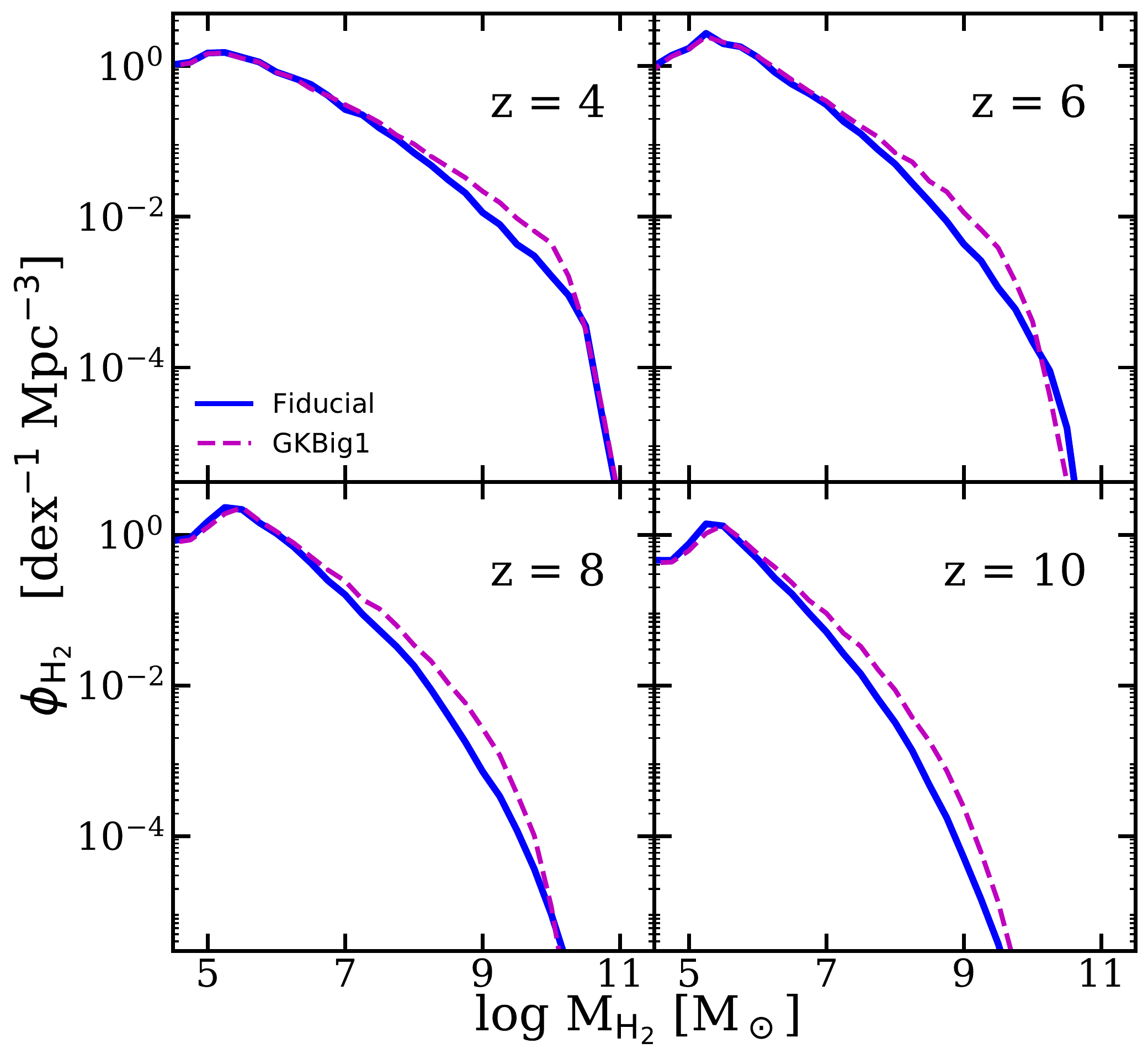}
    \caption{Predicted distribution functions for molecular hydrogen $M_\text{\molh}$ at $z = 4$, 6, 8, and 10. The blue solid line shows the results of the GK-Big2 (fiducial) model and the purple dashed line shows the GK-Big1 model. These models yielded very different predictions for the SMF, but predict similar $M_\text{\molh}$ mass functions.}
    \label{fig:H2MF}
\end{figure}

In this subsection, we show and compare the one-point distribution functions of cold gas and molecular gas masses among our three star formation models. Throughout this work, cold gas mass is defined as $M_\text{cold} = 1.4(M_\text{\ion{H}{I}}+M_\text{\ion{H}{II}}+M_\text{\molh})$, where the factor of $1.4$ is the correction for the mass of \ion{He}{}. In fig. \ref{fig:CGMF}, we compare the cold gas mass functions (CGMF) for galaxies across the three SF models. Note that these models are calibrated to match the SMF, gas fraction, and other observables at $z \sim 0$.  Overall, there are relatively more low-$M_\text{cold}$ galaxies in the \molh-based SF models (GK-Big1 and Fiducial) compared to the $M_\text{cold}$-based KS model due to the the more efficient scaling between cold gas to SF in low-mass halos (see fig. 1 in \citetalias{Somerville2015}). Given that the low-$M_*$ and low-SFR galaxy populations are nearly identical as shown in fig. \ref{fig:SMF} and \ref{fig:SFRF}, we can infer that gas depletion has been much slower in low-mass galaxies in \molh-based models. The difference narrows toward low redshift. This is largely due to the adopted metallicity dependent formation efficiency of \molh\ in the GK-based models --- \molh\ formation is less efficient in low-mass galaxies at at high redshifts, because of the lower metallicities in both cases.

\begin{figure}
    \includegraphics[width=\columnwidth]{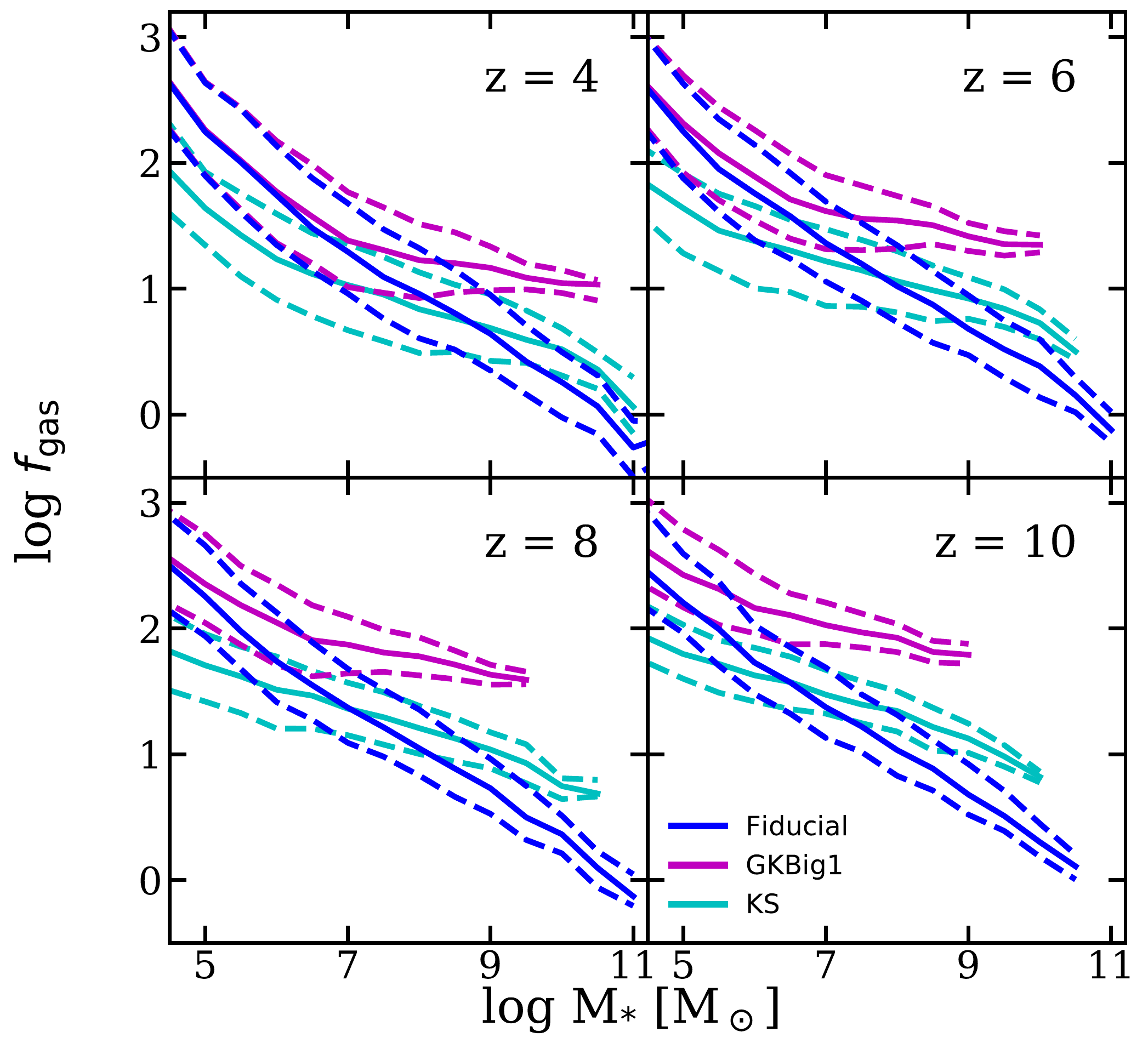}
    \caption{Predicted relationship between stellar mass and cold gas fraction $f_\text{gas} = M_\text{cold}/M_*$ at $z = 4$, 6, 8, and 10. The solid line marks the median and the dashed lines mark the 16th and 84th percentile. The blue lines show the results of GK-Big2 (fiducial) model, the purple dashed lines show the GK-Big1 model, and the cyan dot-dashed lines show the KS model. Varying the star formation recipe yields dramatically different predictions for the cold gas content of galaxies at high redshift.}
    \label{fig:fgas}
\end{figure}

\begin{figure}
    \includegraphics[width=\columnwidth]{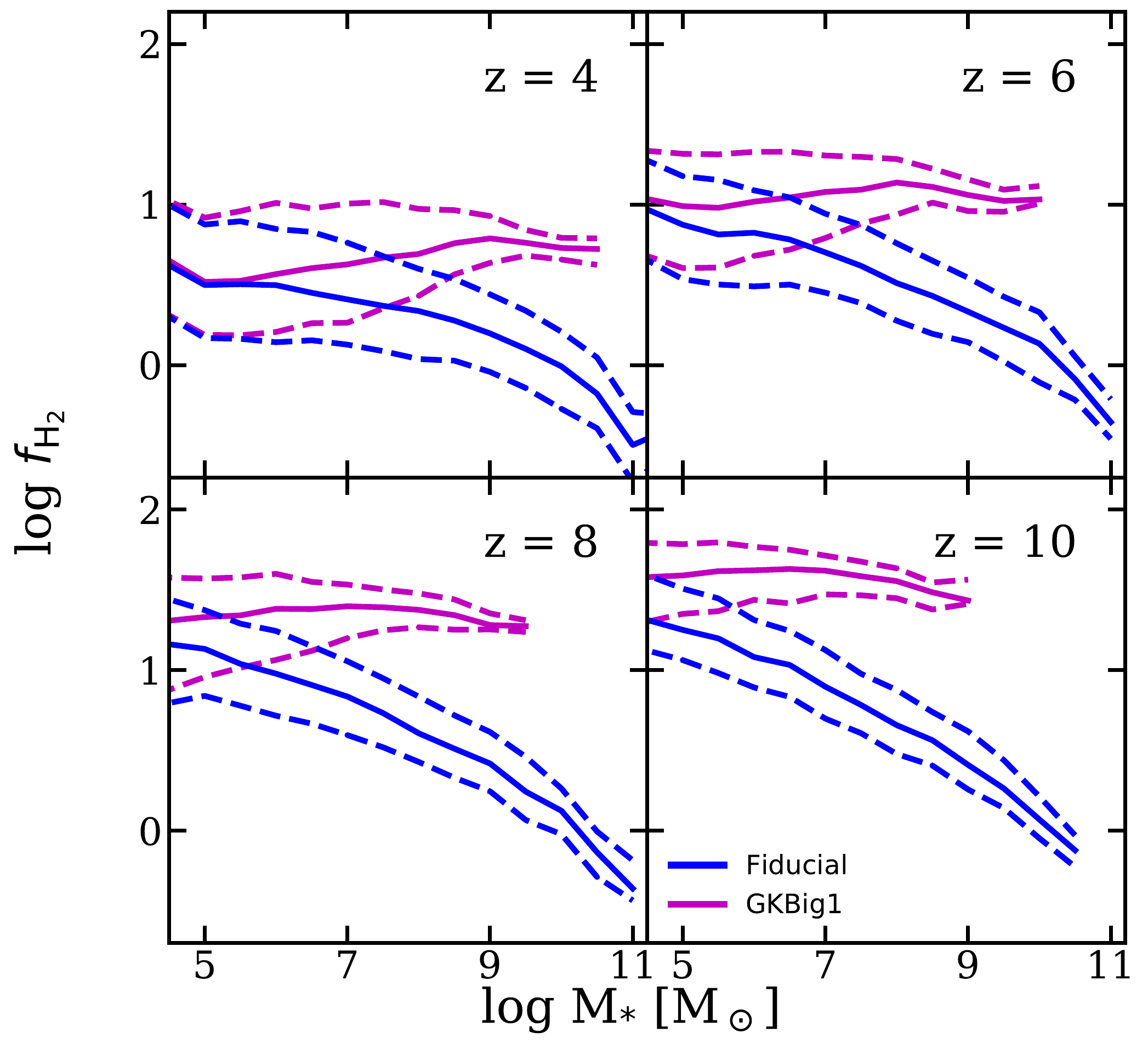}
    \caption{Predicted relationship between stellar mass and molecular gas fraction $f_\text{\molh} = M_\text{\molh}/M_*$ at $z = 4$, 6, 8, and 10. The solid line marks the median and the dashed lines mark the 16th and 84th percentiles. The blue lines show the results of the GK-Big2 (fiducial) model and the purple dashed lines show the GK-Big1 model. The $M_\text{\molh}$ content of galaxies is very sensitive to the star formation recipe. }
    \label{fig:fH2}
\end{figure}

In fig. \ref{fig:H2MF}, we show the \molh\ mass function (\molh MF) and take a closer look at the difference between the Fiducial and the \citetalias{Gnedin2011}-\citetalias{Bigiel2008}1 models. Note that predictions for molecular gas content are only available in models with multiphase gas partitioning (the ones labeled \citetalias{Gnedin2011}). Recall that these two models are very similar, except that the SF relation slope in the fiducial model steepens to $\sim 2$ at higher surface density, as suggested by some observational evidence, while it remains constant in the \citetalias{Gnedin2011}-\citetalias{Bigiel2008}1 model. Thus, the differences in \molh\ mass between these models allow us to indirectly probe the dependence of star formation on \molh\ density. We see that the fiducial model has fewer galaxies with high \molh\ content at all redshifts studied, but the difference appears relatively subtle.

We further investigate the differences in these models by looking into the cold gas fraction, which is defined as $f_\text{gas}\equiv M_\text{cold}/M_*$, shown in fig. \ref{fig:fgas}. For each model, we show the scaling relation between $f_\text{gas}$ and $M_*$ for all three of our SF models, where the solid lines mark the median and the dashed lines mark the 16th and 84th percentile. As mentioned before, the \molh-based models have significantly higher gas fractions especially in low-$M_*$ galaxies, because of the low \molh\ content of these galaxies and resulting low SF efficiency. Similarly, we also show the molecular gas fraction, defined as $f_\text{\molh}\equiv M_\text{\molh}/M_*$, shown in fig. \ref{fig:fH2}. This figure perhaps most clearly illustrates how the fiducial and GK-Big1 models differ. 

As discussed previously, and shown in SPT15, high $M_*$ galaxies have longer \molh\ depletion times (lower efficiencies for converting molecular gas into stars) in the GK-Big1 model than in the fiducial model at high redshift, and the timescale for converting \molh\ into stars is the rate limiting factor for forming stars in these galaxies. Therefore, we see much larger reservoirs of ``leftover'' molecular gas in the GK-Big1 model, also explaining why it fails to produce as many galaxies with large stellar masses at high redshift.

To briefly summarize this entire section, we find that SMFs and SFRFs predicted by our fiducial model agree well with available observations, within the large observational uncertainties. In both cases, the choice of star formation recipe mainly affects the most massive and rapidly star forming galaxies. Varying the efficiency of stellar feedback shifts the location of the low-mass turnover and changes the slope of the low mass end of the SMF. Varying the star formation timescale (normalization of the SF recipe, or gas depletion time) mainly impacts the massive end of the SMF at very high redshifts, where star formation has not yet become self-regulated. The cold gas and \molh\ content of high redshift galaxies is quite sensitive to the details of the star formation recipe, and can help break degeneracies between star formation efficiency and stellar driven wind parameters. For example, the cold gas fraction can discriminate between a model with high star formation efficiency and strong ejective feedback, vs. a model with low star formation efficiency and weaker feedback.

\section{Scaling relations for high-redshift galaxy populations}
\label{sec:scaling}

In this section, we investigate the scaling relations among galaxy properties for the high-redshift populations predicted by our fiducial model that are expected to be detected by \textit{JWST}.

\begin{figure*}
    \includegraphics[width=1.8\columnwidth]{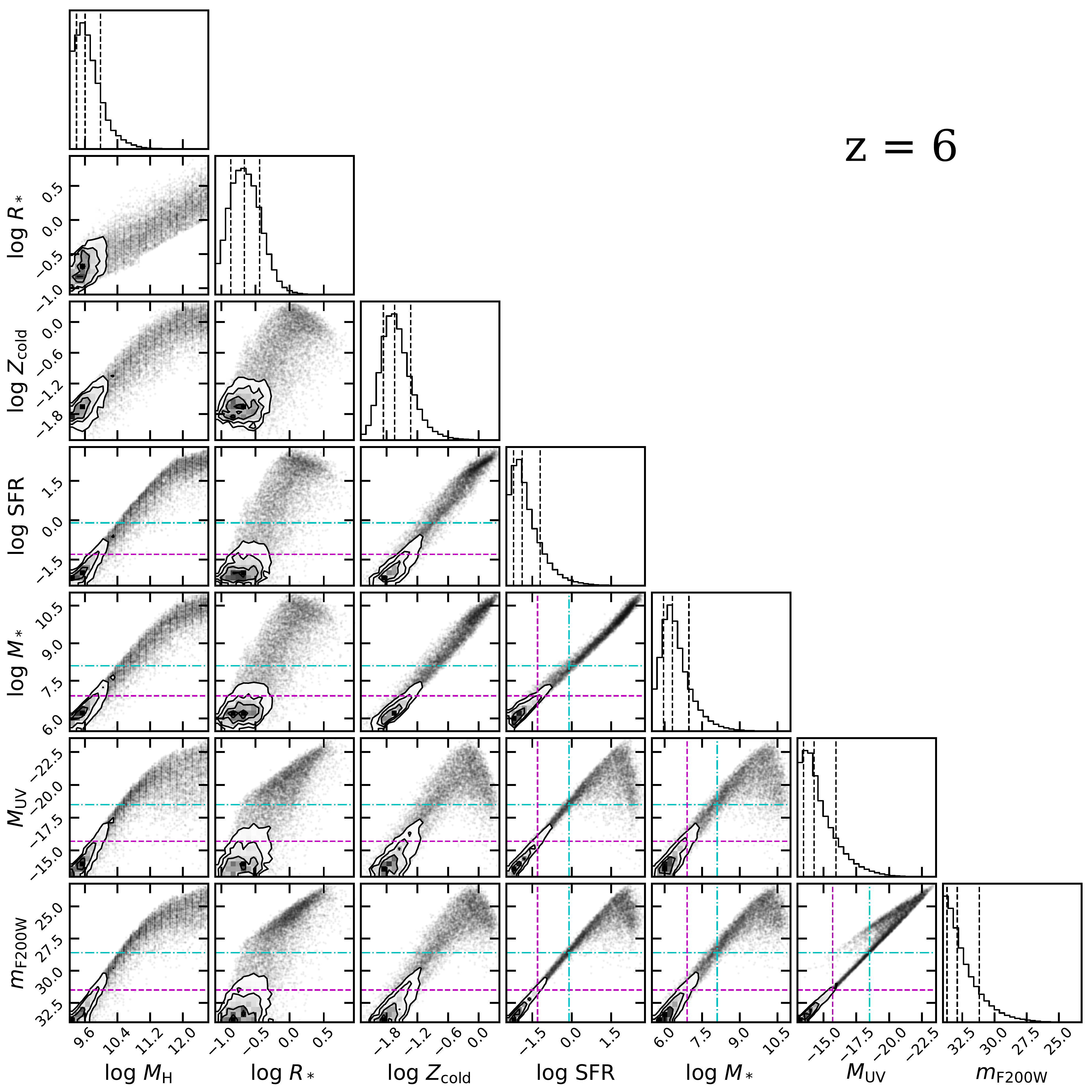}
    \caption{We show relations between halo mass $M_\text{H}$, stellar disk scale radius $R_*$, cold gas phase metallicity $Z_\text{cold}$, SFR, stellar mass $M_*$, rest-frame $M_\text{UV}$ (with dust attenuation), and $m_\text{F200W}$ at $z = 6$ for galaxies above the detection limit of \textit{JWST} lensed surveys ($m_\text{F200W} \lesssim 34.0$) predicted by the fiducial model. All masses are in solar units and $R_*$ is in units of physical kpc. The diagonal panels show histograms for the quantities marked on the corresponding $x$-axis, where the vertical dashed lines mark the 16th, 50th, and 84th percentiles. The off-diagonal panels show the distribution between two properties using a two-dimensional hybrid histogram-scatter plot, which is color-coded for the relative object abundances among the bins. The contours mark the 16th, 50th, and 84th percentiles, while objects falling below the 16th percentile are plotted as individual points. The detection limits for \textit{JWST} wide- and deep-field observations are marked with a cyan dot-dashed line and a purple dashed line, respectively. The values for $M_*$, SFR, and $M_\text{UV}$ corresponding  to these detection limits are marked with matching styles (see Table \ref{table:limit_prop}). The axes of all quantities are oriented such that brighter, more massive, larger objects are to the top or right of the plots, and faint, low-mass, compact objects are to the bottom or left of the plots. }
    \label{fig:corner_z6_trimmed}
\end{figure*}

\begin{figure*}
    \includegraphics[width=1.8\columnwidth]{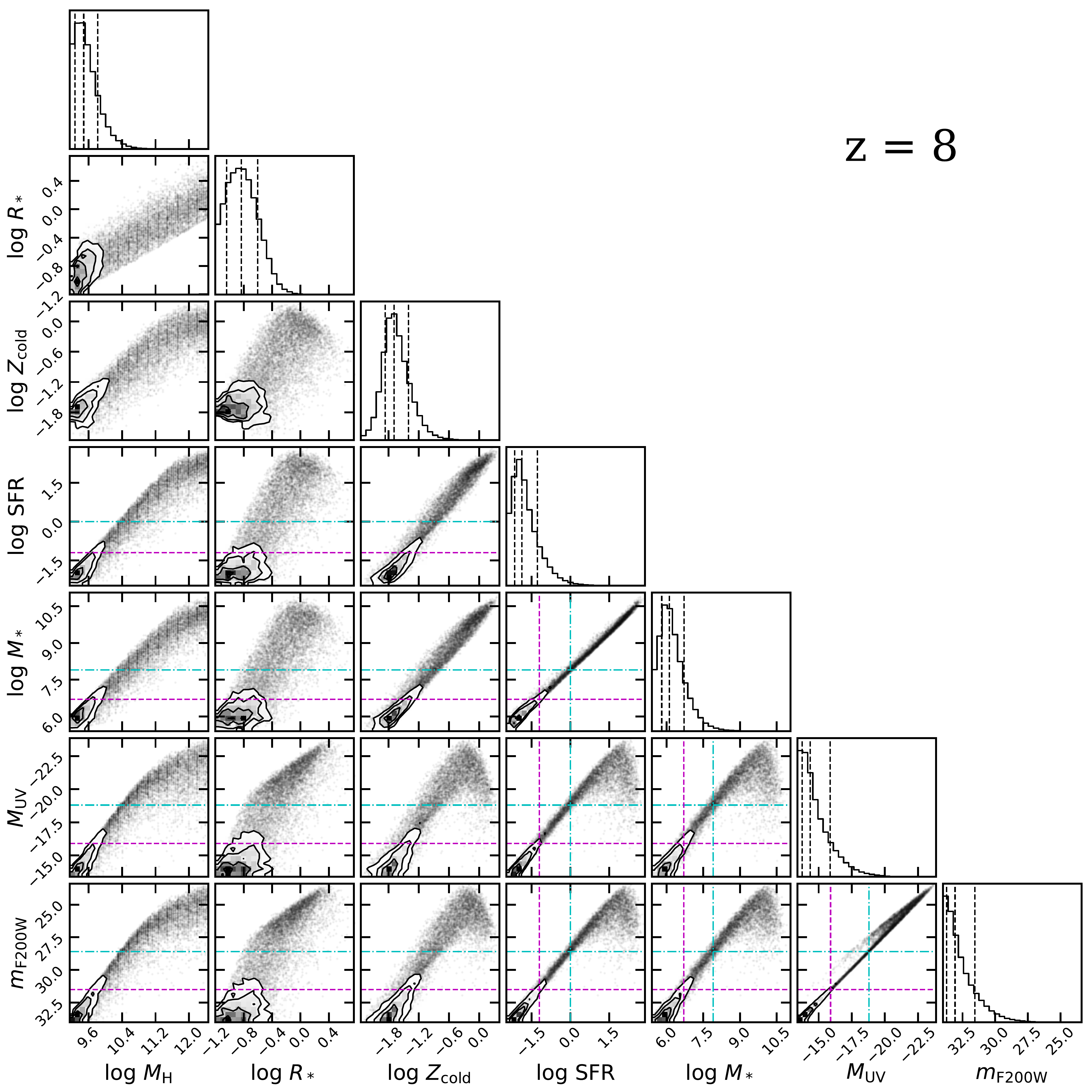}
    \caption{Same as fig. \ref{fig:corner_z6_trimmed} but for galaxies at $z = 8$.}
    \label{fig:corner_z8_trimmed}
\end{figure*}

\begin{figure*}
    \includegraphics[width=1.8\columnwidth]{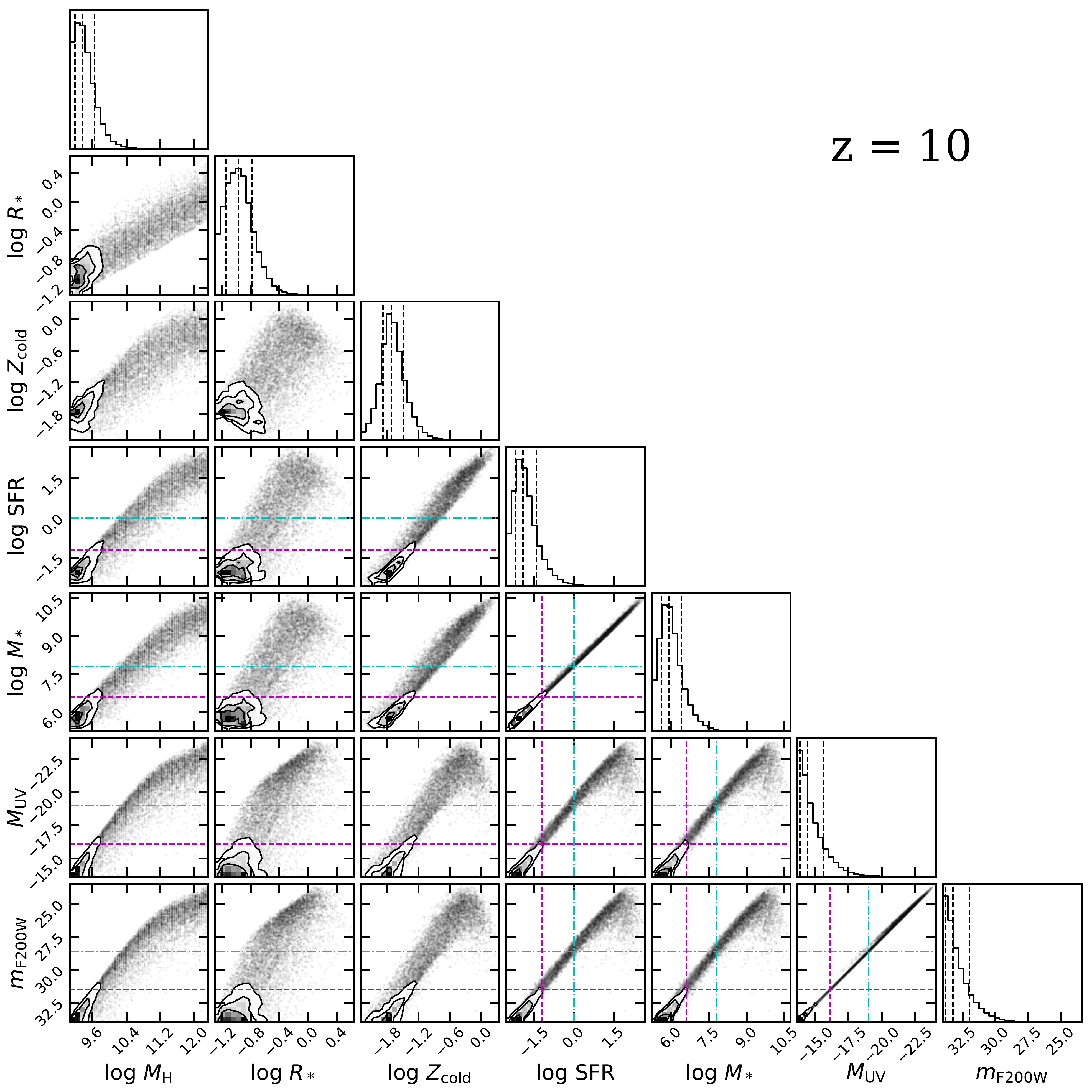}
    \caption{Same as fig. \ref{fig:corner_z6_trimmed} but for galaxies at $z = 10$.}
    \label{fig:corner_z10_trimmed}
\end{figure*}

In fig. \ref{fig:corner_z6_trimmed}, \ref{fig:corner_z8_trimmed}, and \ref{fig:corner_z10_trimmed}, we present the distributions between every permutation among a few selected photometric and physical properties at $z = 6$, 8, and 10, respectively, for galaxies that are detectable given the very optimistic detection limit for lensed surveys ($m_\text{F200W} \lesssim 34.0$). These plots are prepared using the \texttt{corner.py} module provided by \citet{Foreman-Mackey2016}. Plots for other redshifts are omitted to avoid clutter and are made available online. The properties included here are halo mass ($M_\text{H}$), scale radius of the stellar disc ($R_*$), cold gas metallicity ($Z_\text{cold}$), SFR, $M_*$, dust attenuated rest-frame UV luminosity ($M_\text{UV}$), and observed frame IR magnitude in the NIRCam F200W filter $m_\text{F200W}$. The diagonal panels show one-dimensional histograms for the  property labeled on the corresponding  $x$-axis, with the 16th, 50th, and 84th percentiles of the distribution marked by the vertical lines. Each of the off-diagonal panels shows the distribution between two properties using a two-dimensional hybrid histogram-scatter plot, where the 2D histograms are weighted by the halo abundances. Objects with abundances that fall below the 16th percentile are plotted as individual points. We also mark the detection limits assumed for \textit{JWST} wide surveys  ($m_\text{F200W} \sim 28.6$) and deep surveys ($m_\text{F200W} \sim 31.5$) in these plots. Note that the turnovers in the histogram for halo mass and other quantities are not from the resolution limit of our model, but rather due to the intrinsic scatter in galaxy and halo properties at the detection limit.

These diagrams can be used in several ways. One way is for observers to be able to quickly estimate the range of expected physical properties for galaxies with a given observed frame magnitude. To use the plots in this way, one can look across the columns at a given value of $m_\text{F200W}$. Alternatively, simulators or modelers can use them to estimate the expected physical and observable properties of galaxies hosted by halos of a given mass. To use the plot in this way, one can look up the rows at a given value of $M_\text{H}$. In addition to the normalization and slope of the scaling relations between various quantities, the expected scatter in different quantities is also very important. We notice that some relations (such as stellar mass vs. SFR) are very tight, while others (such as all relations with disk size), show almost no correlation. Some relations are linear and monotonic, while others have a break (such as halo mass vs. stellar mass, SFR, and rest UV luminosity). These scaling relations can also be used to guide and aid in the comparison with semi-empirical models.

\begin{table}
    \centering
    \caption{$M_\text{UV}$, $M_*$, and SFR for galaxies at the detection limit of a representative \textit{JWST} wide survey ($m_\text{F200W,lim} = 28.6$) and deep survey ($m_\text{F200W,lim} = 31.5$).}
    \label{table:limit_prop}
    \begin{tabular}{ccccc}
        \hline
        & $M_\text{UV}$ & $\log M_*$ [\Msun]& $\log$ SFR [\Msun\ yr$^{-1}$] \\
        \hline
        $z$ & Wide | Deep & Wide | Deep & Wide | Deep \\ 
        \hline
        4 & -17.03 | -14.40 & 7.92 | 6.75 & -0.73 | -1.82 \\
        5 & -17.76 | -15.10 & 7.99 | 6.83 & -0.42 | -1.53 \\
        6 & -17.97 | -15.32 & 7.92 | 6.76 & -0.32 | -1.43 \\
        7 & -18.11 | -15.47 & 7.84 | 6.68 & -0.27 | -1.38 \\
        8 & -18.23 | -15.56 & 7.78 | 6.61 & -0.21 | -1.35 \\
        9 & -18.34 | -15.65 & 7.69 | 6.53 & -0.22 | -1.36 \\
        10 & -18.32 | -15.68 & 7.57 | 6.44 & -0.27 | -1.39 \\
        \hline
    \end{tabular}
\end{table}

We see fairly tight correlations between $m_\text{F200W}$ and $M_\text{UV}$, SFR, and $M_*$. We track the median of these relations and their redshift evolution, and find that they remain linear and evolve steadily between $z = 4$--10. Fitting functions for selected scaling relations are presented in table \ref{table:scaling_fit} in Appendix \ref{appendix:c}. Note that the correlations are disrupted in the bright, massive populations due to the effect of dust attenuation, which is not accounted for in the fitting. The median values of $M_\text{UV}$, SFR, and $M_*$ specifically corresponding to the wide- and deep-field detection limits across redshifts are then presented in Table \ref{table:limit_prop}. These corresponding values are also marked in fig. \ref{fig:corner_z6_trimmed}, \ref{fig:corner_z8_trimmed}, and \ref{fig:corner_z10_trimmed} with lines of matching color and style.

As illustrated in the last panel of fig. \ref{fig:SMF_JWST} and \ref{fig:SFRF_JWST}, it is interesting to note that the limiting $M_*$ and SFR for a given observed-frame detection limit seems to be evolving only slightly across $z = 4$--10. Intuitively, one would expect a more rapid evolution toward higher $M_*$ and SFR at higher redshift, because of the dimming due to the larger luminosity distances. However, our results demonstrate otherwise. As shown in Table \ref{table:limit_prop}, for galaxies with a certain observed-frame IR magnitude, high-redshift galaxies seem to have higher rest-frame UV luminosities, lower stellar masses, and lower SFRs compared to their low-redshift counterparts. Equivalently, we can say that galaxies at high redshift are intrinsically brighter than their low-redshift counterparts of similar rest-frame UV luminosities, stellar masses, or SFRs (also see fig. \ref{fig:everything_ms}). This effect is apparently nearly sufficient to cancel out the dimming effect.

\section{Redshift evolution of scaling relations}
\label{sec:scaling_z}
\begin{figure}
    \includegraphics[width=\columnwidth]{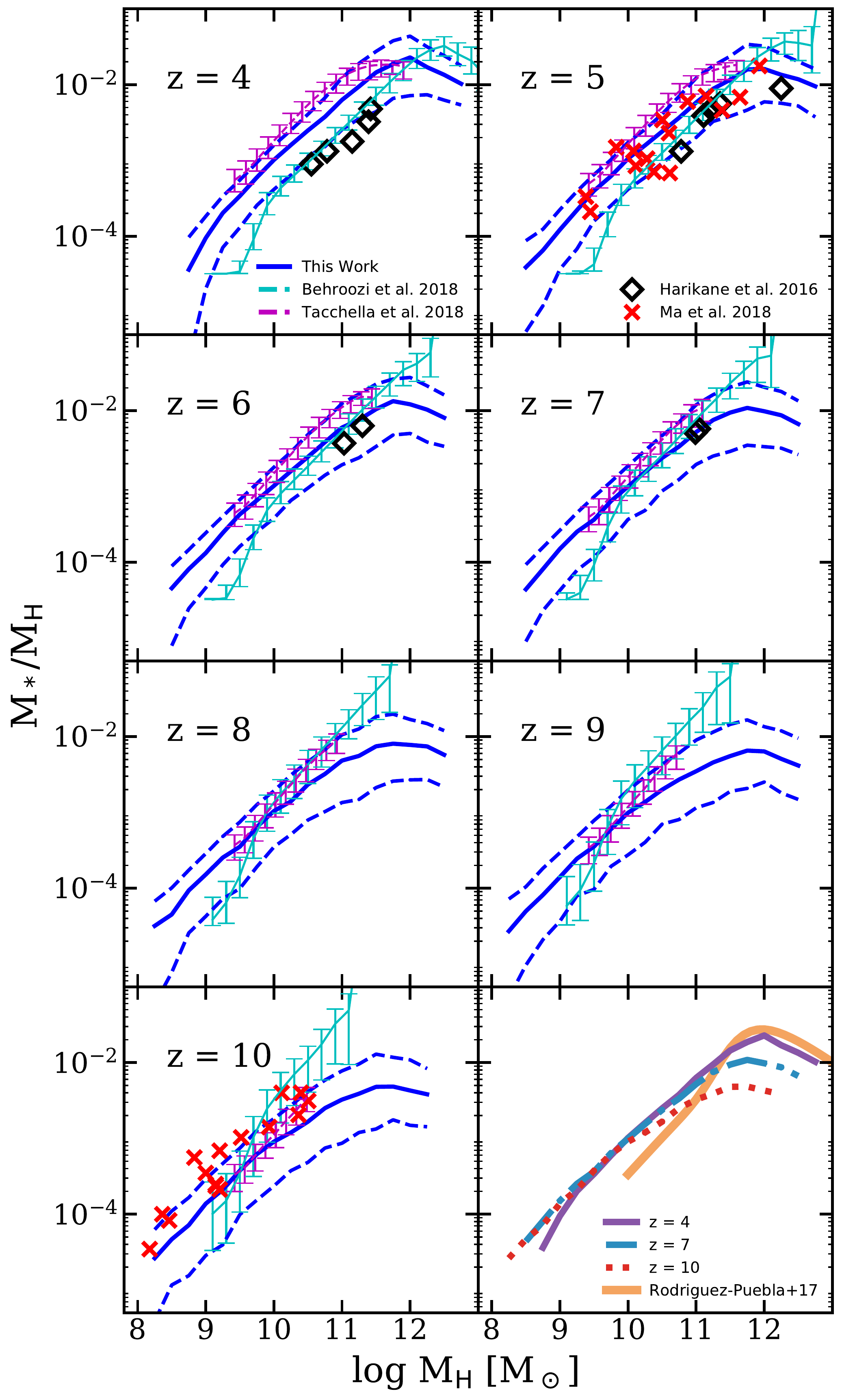}
    \caption{Stellar-to-halo mass ratio (SHMR) predicted by our fiducial model (dark blue). The solid line shows the median, and the dashed lines show the 16th and 84th percentiles. Our results are compared to semi-emperical models from \citet{Behroozi2018} and \citet{Tacchella2018}. The error bars represent the statistical errors in these models. The red cross symbols show predictions from the FIRE-2 simulations \citep{Ma2017} and observational constraints from the clustering analysis by \citet{Harikane2016}. The last panel show an overlay of the SHMR median predicted at $z = 4$, 7, and 10 from our model. We also show abundance matching results from \citet{Rodriguez-Puebla2017} at $z = 0$, which is used in the calibration of our model. }
    \label{fig:SHMR}
\end{figure}

In this section, we focus on the redshift evolution of scaling relations and other physical properties across $z = 4$--10. In fig. \ref{fig:SHMR}, we show the stellar-to-halo mass relation (SHMR) predicted by our fiducial model and compare it to predictions from other models, including semi-empirical model predictions from \citeauthor{Tacchella2018} and \textsc{UniverseMachine}. The Tacchella et al. model assumed a Salpeter IMF, and as above we have applied a correction to the stellar masses throughout this section to make them more appropriately comparable to a Chabrier IMF, used in our work. For our predictions, we show the median and the intrinsic scatter of the distributions (16th and 84th percentiles). For the \citeauthor{Tacchella2018} and \textsc{UniverseMachine} model outputs, the statistical errors are shown. In the last panel we show the median of the relations predicted at $z = 4$, 7, and 10 to highlight the evolution.

Note that \citeauthor{Tacchella2018} assumed a WMAP-7 cosmology, while \textsc{UniverseMachine} assumes Planck 2015 cosmology that is consistent with the one adopted in this work. As shown in fig. \ref{fig:HMF_check_2} in Appendix \ref{appendix:d}, changing from the WMAP-7 to Planck15 cosmology increases the abundance of halos across all masses, and thus effectively shifts the halo occupation function. Thus halos of the same mass must contain galaxies of lower stellar mass in the Planck15 than in WMAP-7 cosmology. It is intriguing that the results from these models are very consistent at $z = 8$ and 9, but then diverge substantially at lower redshift. We speculate that this is due to both the differences in assumed cosmology and in the observational constraints used by these models. As shown in the lower panel of fig. \ref{fig:HMF_check_2}, as we move from the legacy WMAP cosmology to the more recent Planck cosmology, the fractional difference in number density of halos in the mass range of interest is smallest at $z = 10$, and later increases rapidly toward lower redshifts. This effectively changes the underlying halo populations assumed for the predicted SHMRs and ultimately shifts the relation horizontally. Moreover, a change in cosmology leads to different accretion rates for dark matter halos, which both of these models depend on. On the other hand, as shown in fig. \ref{fig:smf_models}, the predicted SMFs from these models seem to be in good agreement at $z \gtrsim 8$ but the predictions become quite different at $z = 4$, with the \citeauthor{Tacchella2018} model predicting $\sim0.5$ dex more galaxies than \textsc{UniverseMachine}. This discrepancy is a result of the differences in observational constraints at low redshift used for calibration and the assumptions made in these models (see explanation in \S\ref{sec:models}).

\begin{figure}
    \includegraphics[width=\columnwidth]{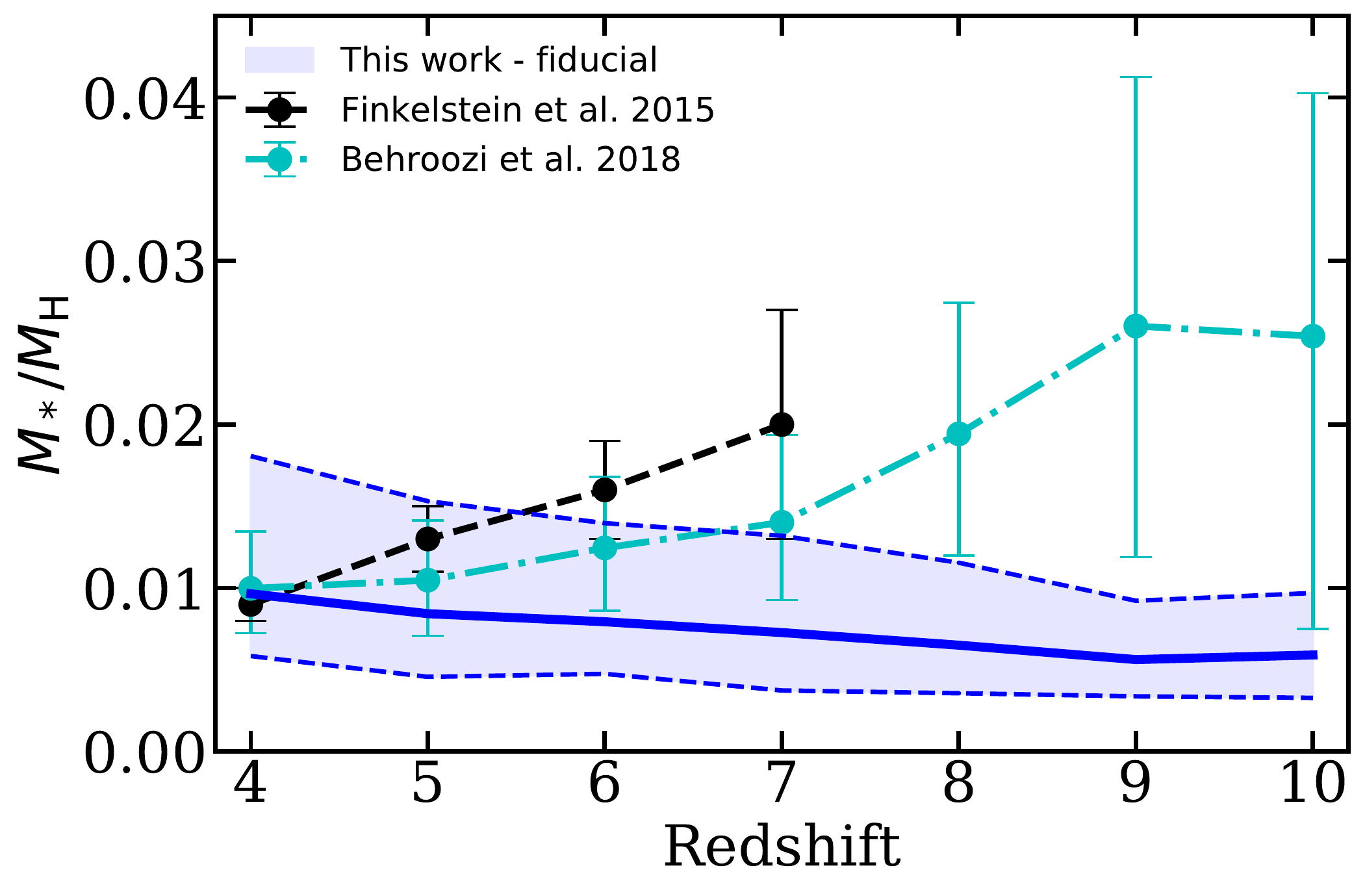}
    \caption{SHMR as a function of redshift predicted by our fiducial model for galaxies with $M_\text{UV} = -21 \pm 0.25$ compared to observational estimates from \citet{Finkelstein2015a} and the empirical model \textsc{UniverseMachine} \citep{Behroozi2018}. The blue solid line shows the median from our model, and the dashed lines mark the 16th and 84th percentiles. The error bars for \citeauthor{Finkelstein2015a} and \textsc{UniverseMachine} represent observational uncertainties and statistical errors. Our models predict that the SHMR decreases slightly with increasing redshift for these galaxies, while the observational estimates and the \textsc{UniverseMachine} predict an increase of this quantity back in time. }
    \label{fig:SHMR_evo}
\end{figure}

In addition to the empirical models, we compare our results to the FIRE-2 simulations \citep{Ma2017} and to the clustering analysis by \citet{Harikane2016}. \citeauthor{Ma2017} reported both $M_*$ and $M_\text{h}$ for 15 galaxies at the end of the simulations ($z = 5$) and their progenitor halos at $z = 10$. The \citeauthor{Harikane2016} analysis is based on 10,381 Lyman break galaxies (LBGs) at $z \sim 4$--7 identified in various legacy surveys. We also show the evolution of the SHMR at $z = 4$, 7, and 10, and compared that to abundance matching result at $z = 0$ from \citet{Rodriguez-Puebla2017}. Our predictions for the SMHM relation are remarkably similar to the FIRE results. They are consistent with the \citeauthor{Harikane2016} results at $z\sim 6$ and $z\sim 7$, but diverge by an increasing amount at $z\sim 5$ and $z\sim 4$. It is interesting that the clustering-based \citeauthor{Harikane2016} results are more consistent with \textsc{UniverseMachine}.

\begin{figure}
    \includegraphics[width=\columnwidth]{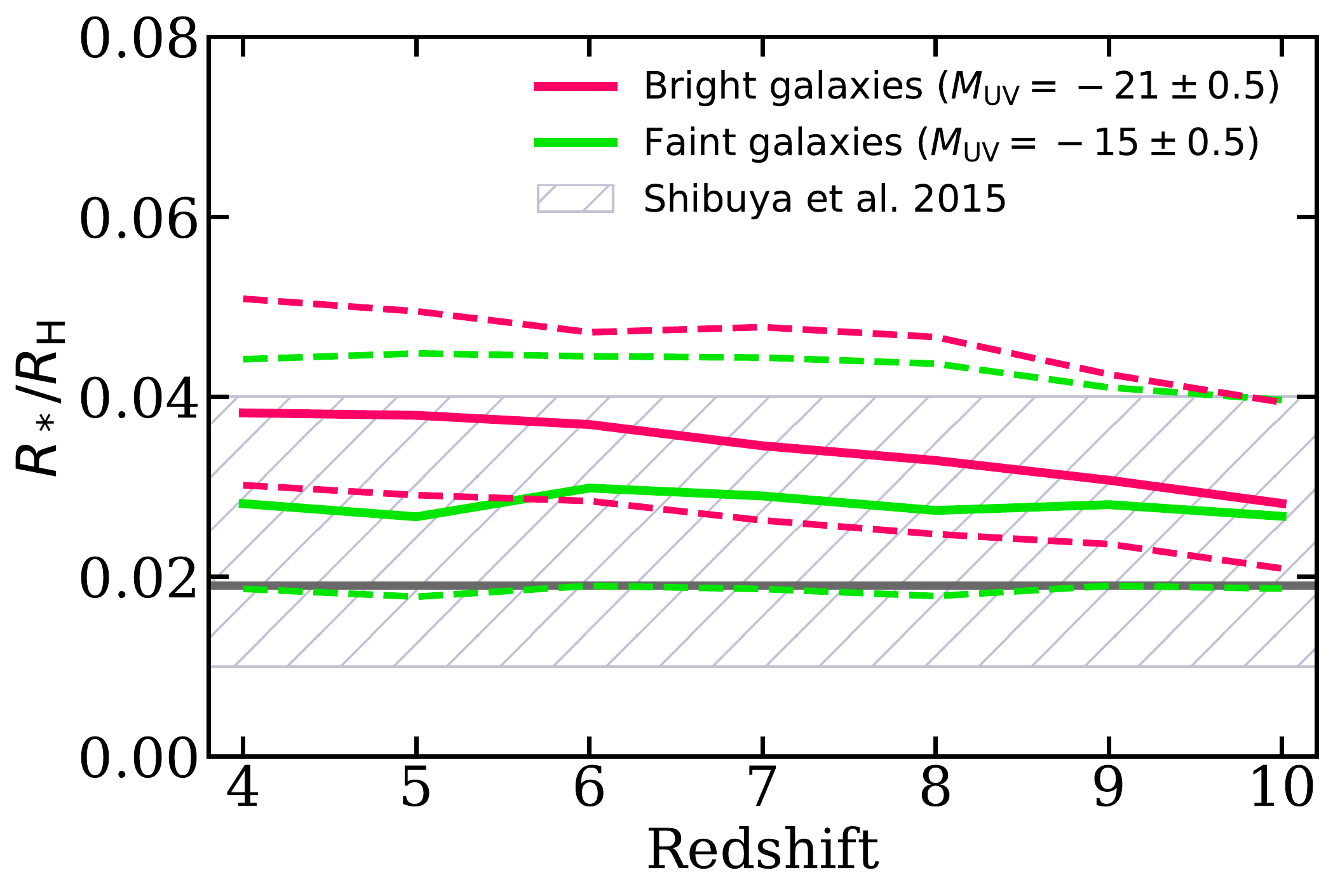}
    \caption{The redshift evolution of the stellar-to-halo size ratio predicted by our model, where the results for bright ($M_\text{UV} = -21\pm0.5$, purple) and faint ($M_\text{UV} = -15\pm0.5$, blue) galaxies are presented separately. The error bars represent the intrinsic scatter of the properties, where the upper and lower limits mark the 84th and 16th percentile, respectively. The gray band approximates the abundance matching results presented by \citep{Shibuya2015} ($z \sim 4$--8). The radius shown for our model is the 3D half-stellar mass radius, which is expected to be slightly larger than the projected, rest-UV effective radius measured by observations. It is intriguing that the relationship predicted by our model is in good agreement with the observational constraints out to $z\sim8$.}
    \label{fig:shrr_evo}
\end{figure}

\begin{figure*}
    \includegraphics[width=2\columnwidth]{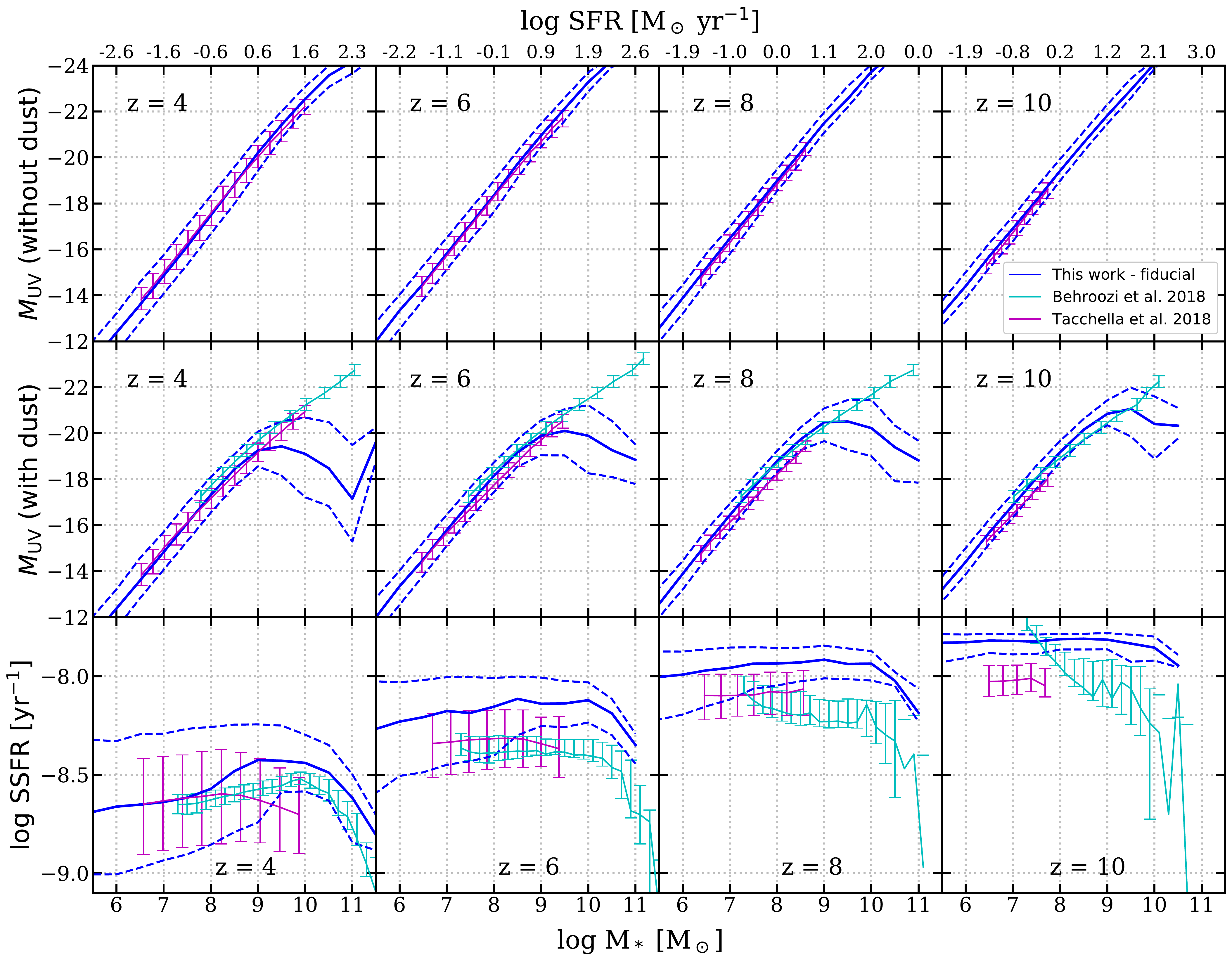}
    \caption{The redshift evolution of $M_\text{UV}$ (top row: unattenuated; middle row: with dust attenuation) and sSFR predicted by our fiducial model (dark blue), where  the 16th, 50th, and 84th percentiles are shown. Predictions from \textsc{UniverseMachine} \citep{Behroozi2018} and \citet{Tacchella2018} are also shown in cyan and purple, respectively. The error bars represent the statistical errors on these models. This figure illustrates that there are significant differences in the assumed and predicted scaling relations both between different semi-empirical models and between semi-analytic and semi-empirical models.
    }
    \label{fig:everything_ms}
\end{figure*}

In fig. \ref{fig:SHMR_evo}, we compare the evolution of the SHMR for a specific range of observed galaxies with $M_\text{UV} = -21 \pm 0.25$ as presented in \citet{Finkelstein2015a}. The halo masses for these observed galaxies are estimated using abundance matching with halo populations assuming the WMAP-7 cosmology. Similarly, galaxies predicted by our model and the \textsc{UniverseMachine} that satisfy the same criteria are selected for comparison. It is quite intriguing that abundance matching and empirical models indicate an increasing trend out to higher redshift, while our physically motivated model predicts the opposite. This highlights an interesting open question that will be addressed by future \textit{JWST} observations.

Similarly, in fig. \ref{fig:shrr_evo}, we show the evolution of the ratio between the stellar disc radius and the virial radius of the host halo (stellar-to-halo radius ratio, SRHR) for the bright ($M_\text{UV} = -21.0 \pm 0.5$) and the faint ($M_\text{UV} = -15.0 \pm 0.5$) populations predicted by our models. These predictions are compared to observations between $z \sim 4$--8 from \citet{Shibuya2015}. Instead of showing individual measurements, we show the approximate median and uncertainties $R_*/R_\text{H} \approx 0.019^{+0.21}_{-0.09}$. The radius shown for our model is the 3D half-stellar mass radius, which is expected to be slightly larger than the projected, rest-UV effective radius measured by observations. The rather simple relation adopted in our model is in good agreement with the observations out to $z\sim8$. This highlights another interesting science question that will be elucidated by \textit{JWST} observations.

In fig. \ref{fig:everything_ms} we show a side by side comparison of the $M_*-M_\text{UV}$ and $M_*-$sSFR relation at $z = 4$, 6, 8, and 10, compared with the relations obtained from or assumed in \citeauthor{Tacchella2018} and \textsc{UniverseMachine} \citep{Behroozi2018}. Using the linear fit we found for the $M_*-$SFR relation, we also label the $x$-axis with the value of SFR corresponding to the given $M_*$. We find that sSFR evolves quite rapidly as a function of redshift. These models agree quite well at $z = 4$ but then diverge at high redshifts. The pronounced differences at high stellar masses arise from our different treatments of dust attenuation. We can also see the different ranges in halo mass spanned by the \citeauthor{Tacchella2018} model and \textsc{UniverseMachine} models, which are due to their use of numerical $N$-body simulations, which suffer from limited mass resolution and volume. Our predictions span the largest dynamical range, because of our use of analytic halo mass functions calibrated to multiple simulations with varying resolution and volume.

\section{Discussion}
\label{sec:discussion}

In this section, we discuss some caveats and uncertainties in our modeling, discuss our results in the context of other results in the literature, and present an outlook for future observations beyond \textit{JWST}.

\subsection{Interplay between galaxy formation and cosmology}
The details of when galaxies form and how quickly they evolve are somewhat sensitive to the adopted cosmology and primordial power spectrum. The estimated values of the cosmological parameters have evolved significantly throughout the past few years. For instance, the matter density parameter $\Omega_m$ measured by the \citet{Planck2014, Planck2016} decreased significantly from previous measurement by WMAP \citep{Komatsu2009, Komatsu2011, Hinshaw2013}. This has non-trivial effects on the dark matter halo demographics (see fig. \ref{fig:HMF_check_2} in Appendix \ref{appendix:d}). In this work, we have updated the cosmological parameters adopted in our model to the more up-to-date values reported by the \citeauthor{Planck2016} in 2016, as compared to the WMAP-5 values used in previous work with these models. As $\Omega_m$ increases, the abundance of dark matter halos found in cosmological $N$-body simulations increases across all mass scales. In order to continue to match the same observed galaxy abundances, the relationship between galaxy luminosity or stellar mass and halo mass has to shift, such that the stellar mass occupying a halo of a given mass must decrease. In physically based models of galaxy formation such as our SAMs, this is accomplished by changing the efficiency of stellar feedback and/or the star formation efficiency. We provide details on the changes in parameter values that were required in our models in order to recalibrate to the Planck cosmology in \citetalias{Yung2019}. It would seem to provide indirect support for the $\Lambda$CDM cosmology and hierarchical structure formation paradigm that implementing baryonic processes in a physically motivated manner is able to qualitatively  reproduce the observed assembly of stellar mass from $z\sim 8$ to the present day.

\subsection{Halo mass functions and merger histories}
Understanding how gravitationally bound dark matter halos form and evolve over cosmic time provides the scaffolding for any cosmologically grounded model of galaxy formation. However, predictions for the global galaxy population from the reionization epoch to the present day require a very large dynamic range (at least seven orders of magnitude in halo mass). Pure dark matter plus gravity $N$-body simulations are commonly used to extract DM halo properties, abundances, and merger trees. However, there is currently no publically available, self-consistent and comprehensive suite of $N$-body simulations that is consistent with state-of-the-art constraints on the cosmological parameters and that fully spans the required dynamic range in halo mass, resolution, and volume.  Furthermore, most existing $N$-body simulations store only a few outputs at very early times, so that the resulting merger histories are far too coarse at high redshift. Additionally, halo finders and merger tree building algorithms have generally not been tested at these redshifts. As a result, we have made use of fitting functions calibrated to available simulation results (see \citetalias{Yung2019}, Appendix \ref{appendix:d}), combined with merger trees created using a method based on the Extended Press Schechter (EPS) formalism. Although this approach is highly flexible and computationally efficient, allowing us to provide predictions of galaxy properties over the largest range of mass and redshift of any study yet published of which we are aware, the constructed ensembles of merger trees are not in perfect agreement with the results from $N$-body. We have tested our models by running them within merger trees extracted directly from $N$-body simulations, and find good agreement. The EPS method has the additional drawback that it does not predict spatial information or capture any dependences of halo properties and formation history on larger scale environment (which are known to exist in full $N$-body simulations).

\subsection{Uncertainties in modeling of baryonic processes}
In \citetalias{Yung2019} of this series, we showed that a semi-analytic model of galaxy formation that was calibrated only with observations at $z=0$ makes predictions for rest UV luminosity functions at $z\sim 4$--10 that are in remarkably good agreement with available observational estimates. In this Paper, we have similarly shown that the same model also reproduces observational estimates of the stellar mass function and star formation rate function at these epochs (within the rather large observational uncertainties). A significant uncertainty in our modeling of galaxy luminosities, particularly in the rest UV, is our treatment of attenuation by dust, which is very simple and somewhat ad hoc. In \citetalias{Yung2019}, we showed that the dust attenuations adopted in our model are consistent with observational measurements of the UV slope in high-redshift galaxies, which we found encouraging. However, these measurements have very large error bars, and in addition UV slope constrains only reddening and the relationship between attenuation and reddening is not unique or well-constrained, especially at these redshifts. The finding that our models also produce stellar mass and SFR distributions that are consistent with those estimated from observations is an additional consistency check, indicating that the stellar populations in high redshift galaxies in our models are apparently similar to those inferred from fitting the multi-wavelength SEDs of real high redshift galaxies. We emphasize that the median relation, as well as the scatter in, the relationship between stellar mass and rest-UV stellar mass-to-light ratio at $z \gtrsim 4$ is still highly uncertain, and there are significant differences in the relations adopted in the literature. This is an important component in many semi-empirical models. \textit{JWST} will greatly improve the constraints on this critical relationship. Furthermore, semi-analytic models are beginning to model the creation and destruction of dust in galaxies self-consistently \citep*{Popping2017}, which will allow for more physically robust dust modeling in future works.

One of the most striking aspects of our results, though we are not the first to demonstrate this, is that a relatively simple model qualitatively reproduces many of the fundamental properties of galaxies (such as stellar mass functions) from $z = 0$ to $z \sim 10$, without any explicit tuning to high redshift observations or introduction of ad hoc redshift dependence in the recipes. Given the very different conditions at $z \sim 6$--10 compared to the present day Universe, it would not have been at all surprising if our current simple phenomenological parameterizations of physical processes had broken down badly. It is also interesting to note that certain models have already been strongly ruled out by this comparison --- our work, as shown in fig. \ref{fig:SMF} and \ref{fig:SFRF} (see also \S3 in \citetalias{Yung2019}), convincingly demonstrates that in order to produce sufficient numbers of massive/luminous, high redshift galaxies, it is necessary to adopt a star formation relation in which SFR density scales super-linearly with molecular gas surface density in dense gas. Because high redshift galaxies tend to be more compact and have more gas at high surface density, this leads to an effective decrease in the gas depletion time (increase in star formation efficiency) at high redshift. This appears to be in qualitative agreement with existing observations of cold gas at high redshift \citep{Obreschkow2009a, Dutton2010, Dutton2012, Saintonge2013, Decarli2016, Schinnerer2016, Krogager2018, Tacconi2018}, but future observations will provide improved constraints.

Although the fundamental physics operating in our Universe should remain unchanged across space and time, changes in the prevalent physical conditions could lead to effective evolution in redshift or cosmic time for the processes that shape galaxies.  For example, processes that involve interaction with the cosmic environment, such as dark matter halo mergers, photoionization squelching, or cosmological accretion, depend on the background density or temperature, which is redshift dependent. Local processes, such as AGN and stellar feedback, are generally parameterized as a function of a galaxy properties that evolve across redshift and thus gain de facto dependency on redshift. Since each of these processes operates under different conditions and depends on a different set of properties, in principle one might be able to break certain degeneracies by studying a wide range of galaxy types over different snapshots in cosmic time. However, the baryonic processes in SAMs are parameterized in an extremely simple and phenomenological manner, and it is unclear whether these parameterizations will properly capture these multivariate correlations. Further close comparisons of SAM predictions with those from numerical hydrodynamic simulations are important to validate this approach.

There are also physical processes that may be important at extreme redshifts that are not included in our models. For example, our models do not directly model Pop III stars, nor metal enrichment by these objects, but instead assume that all top-level halos are polluted up to a metallicity floor $Z_\text{pre-enrich} = 10^{-3} Z_\odot$. Our models do not include self-consistent modeling of photo-ionization feedback (`squelching') by a meta-galactic ionizing background, but instead assume that reionization occurs everywhere in the Universe at a fixed redshift of $z\sim 8$. As discussed in \citetalias{Yung2019}, in some past works, squelching was thought to have a significant impact on galaxy formation in halos up to masses of $\sim 10^{10}$ \Msun\ (e.g. \citealt{Efstathiou1992}; \citealt*{Bullock2000}; \citealt{Gnedin2000, Somerville2002}). However, in more recent studies, the halo mass where squelching has a significant impact has dropped to much lower masses.  In this work, we adopted the characteristic mass fitting function from \citet{Okamoto2008} and found that squelching has a negligible effect on observable galaxies in the mass and redshift range that we studied. This is in agreement with results from the Cosmic Reionization On Computers \citep[CROC,][]{Gnedin2014a} simulations, but not with the Cosmic Dawn simulation \citep[CoDa,][]{Ocvirk2016}, which found that the photoionizing background has a strong effect on halos with a mass at $z \sim 3$ of $M_\text{H} \lesssim 10^{10}$\Msun. The interplay between photo-ionization feedback and other feedback processes such as stellar feedback is extremely complex and must be further investigated using simulations with self-consistent radiative transfer \citep[e.g.][]{Finlator2011,Finlator2018}.

Another set of processes that is highly uncertain in our models is the seeding and growth of, and feedback from, supermassive black holes. In our current models, we do not include the radiation from AGN in the galaxy SED that we compute, nor do we believe our current treatment of AGN feedback at high redshift to be realistic. This could have an important effect on galaxy properties and reionization, and will be the topic of future work.

\subsection{Our results in the context of other model predictions}
We have performed a fairly comprehensive comparison of our predictions for galaxy stellar mass functions and SFR functions at $z\sim 4$--10 with available predictions from the literature based on the three major existing techniques: semi-analytic models, numerical hydrodynamic simulations, and semi-empirical and empirical models. Somewhat surprisingly, we find that predictions from semi-analytic models and numerical hydrodynamic simulations from several different groups are in generally very good agreement for these basic quantities. This is surprising because these calculations have been done using different codes, different sub-grid treatments of physical processes, and at different resolutions. We see a general consistency among models based on \textit{a priori} modeling of physical processes within a $\Lambda$CDM cosmological framework. Semi-empirical and purely empirical models show larger dispersions, unsurprisingly, in the regions where observational constraints are currently unavailable. This also seems to be encouraging news for efforts to use physically based models for forecasting and planning for future observations.

One of the fundamental questions in galaxy formation and cosmology is how galaxy properties are related to the underlying dark-matter dominated mass distribution. A simplified form of this relationship is often presented in terms of the relationship between galaxy stellar mass and the mass of its host dark matter halo (stellar-mass-halo-mass relation). Our models directly predict this relationship, as well as the dispersion in it, and we find the interesting result that our models predict almost no evolution in the median $M_*/M_\text{H}$ from $z\sim 10$ to $z\sim 4$ for low-mass halos ($M_\text{H} \lesssim 10^{11}$\Msun), while they predict almost an order of magnitude increase over this interval in the most massive halos of these epochs ($M_\text{H} \sim 10^{12}$\Msun).

We compared our predictions for $M_*/M_{\rm H}$ with the results of semi-empirical models from \citeauthor{Tacchella2018} and \textsc{UniverseMachine} at high redshifts. We find significant differences both \emph{between} these two published semi-empirical models and our model predictions. Some of these differences may be due to differences in the assumed underlying cosmology and the observations used to derive the semi-empirical models. Perhaps most strikingly, both \citeauthor{Tacchella2018} and \textsc{UniverseMachine} predict that $M_*/M_\text{H}$ continues to monotonically increase with increasing halo mass at $z \gtrsim 6$, while our models predict that $M_*/M_\text{H}$ turns over at high halo masses. These massive halos are exceedingly rare, and no numerical simulations that we are aware of have investigated this.

Our results are in tension with existing observational estimates of the evolution of $M_*/M_\text{H}$ over the interval $z\sim 4$--7 from \citet{Finkelstein2015a} for a stacked sample of luminous galaxies ($M_{\rm UV} \sim -21$), however, we again emphasize the currently very large uncertainties on these observational estimates. \textit{JWST} is unlikely to place strong constraints on the abundances of these extremely luminous high redshift galaxies, due to its small field of view and limited lifetime, but future wide deep surveys with instruments such as WFIRST can be anticipated to do so. \textit{JWST} will however be able to obtain improved estimates of the redshift and stellar populations in luminous high redshift galaxy candidates that have already been discovered through \textit{HST} and Spitzer.

If it is really true that $M_*/M_\text{H}$ remains essentially constant from $z\sim 4$--10 in the halos that will dominate the populations observed by \textit{JWST} (as our models predict), this has an interesting implication. It implies that the build-up in galaxy number density over this period is driven by the evolution in the dark matter halo mass function. Therefore, if we can somehow observationally constrain this $M_*/M_\text{H}$ relationship, \textit{JWST} observations of high redshift galaxies could provide interesting constraints on cosmology. Another interesting point is that, as can be clearly seen in Fig. \ref{fig:SMF_JWST} (or see also Fig. 13 in \citetalias{Yung2019}), the expected turnover in the stellar mass or luminosity function from baryonic processes is well below the anticipated sensitvity of \textit{JWST} even for lensed fields. Therefore, if a turnover or cutoff at faint magnitudes is seen, it could be a sign of a cutoff in the small scale power spectrum, such as that expected in certain exotic varieties of dark matter.

\subsection{Outlook for future observations with \textit{JWST} and beyond}
One of the main focuses of this work is to establish the connections between the predicted rest-frame UV luminosities and observed-frame IR magnitudes for high-redshift galaxies (presented in \citetalias{Yung2019}) to their intrinsic physical properties. From the wide range of physical properties predicted by our SAM, we have selected a few that are of the greatest interest, including stellar mass $M_*$, SFR, cold gas phase metallicity $Z_\text{cold}$, and stellar radial size $R_*$. A comprehensive view of the correlations among these properties along with $m_\text{F200W}$, $M_\text{UV}$, and halo mass $M_\text{H}$ have been shown in fig. \ref{fig:corner_z6_trimmed}, \ref{fig:corner_z8_trimmed}, and \ref{fig:corner_z10_trimmed}. These corner plots can be used as lookup tables to facilitate both the planning of observations and simulations. For instance, simulators can use the first column to quickly estimate the range of galaxy properties expected for some given halo mass, and conversely, observers can use the bottom row to estimated the physical properties of an observed population given an object's observed-frame IR magnitude.

We applied selection criteria for representative `wide', `deep' and `lensed' \textit{JWST} surveys to our model predictions, in order to predict the range of stellar masses and SFR that will be probed.  By combining these `sensitivity' functions with the experiments that we carried out in \S\ref{sec:properties}, in which we varied the parameters controlling various physical processes in the models, we can see what kind of survey design characteristics will be necessary to probe different physical processes that are currently highly uncertain in galaxy formation models. For example, we can clearly see that the planned wide and deep surveys with \textit{JWST} will help to constrain stellar feedback processes at high redshift. Constraining the efficiency of star formation at redshifts greater than six will likely require larger area surveys than will be feasible with \textit{JWST}.

Apart from the planned \textit{JWST} GTO and ERS programs mentioned above, there will be plenty of other observational opportunities to probe the very early Universe with upcoming facilities, including the space-based Euclid \citep{Racca2016} and Wide-Field Infrared Survey Telescope \citep[WFIRST,][]{Spergel2015}, as well as the ground-based Large Synoptic Survey Telescope \citep{LSST2017}. Future deep Atacama Large Millimeter Array (ALMA) surveys will be able to put constraints on the cold gas mass in $z \sim 3$--7 galaxies via CO and dust continuum observations, as well as to probe the ISM conditions through fine-structure lines.

Beyond individual source detection, intensity mapping is a new technique being developed to indirectly constrain the high-redshift galaxy population (\citealt{Visbal2010}; \citealt*{Visbal2011}; \citealt{Kovetz2017}) over large areas of the sky. Numerous on-going intensity mapping experiments for \ion{H}{I}, CO, \ion{C}{II}, and Ly$\alpha$ are planned or underway, including BINGO \citep{Battye2013}, CHIME \citep{Bandura2014}, EXCLAIM \citep{Padmanabhan2018}, HERA \citep{DeBoer2017}, HIRAX \citep{Newburgh2016}, Tianlai \citep{Chen2012}, LOFAR \citep{Patil2017}, MeerKat \citep{Pourtsidou2016, Santos2017}, CONCERTO \citep*{Serra2016}, PAPER \citep{Parsons2010}. The development of efficient and robust modeling techniques to interpret results from all of these upcoming experiments will be critical to realize their full scientific potential.

\section{Summary and Conclusions}
\label{sec:snc}
In this work, we presented predictions from semi-analytic models for  physical properties of galaxies at $z = 4$--10, an epoch that will be probed by upcoming observations with \textit{JWST}. With EPS-based merger trees, we can efficiently sample halos over a wide mass range, ranging from the ones near the atomic cooling limit to the most massive ones at a given redshift, unlike numerical simulations which are much more limited in the dynamic range that can be simulated. We showed distribution functions for statistical properties such as stellar mass and SFR functions, and investigated how varying the parameters in recipes for the physical processes that shape these galaxy populations affect these results. In addition, we provide predictions of the scaling relations between physical properties that are directly predicted by our simulations, as well as between intrinsic and observable properties. We provide predictions for the properties of the galaxy populations that will be probed by representative wide, deep, and lensed \textit{JWST} surveys.

We have shown the one-point distribution functions for stellar mass, star formation rate, cold gas mass, and molecular gas mass between $z = 4$--10 predicted by our semi-analytic model. We have also studied and quantified the impact of uncertainties in our parameterizations of key processes such as star formation efficiency and the mass loading of stellar driven winds. The free parameters in our model are calibrated once to a subset of observational constraints at $z\sim0$ and are not retuned to match observations at higher redshifts. Thus, it is encouraging that the predictions produced by our fiducial model are in good agreement with observational constraints for SMFs and SFRFs up to $z \sim 8$, although the uncertainties on the observational estimates of these quantities are currently quite large. This suggests that the approach used to model physical processes in galaxy formation models, which has been quite successful at lower redshift, is not failing badly even at these very early epochs. We also compare the predictions for SMF and SFRFs from our semi-analytic approach with those from numerical hydrodynamic simulations and semi-empirical models. We find reasonable agreement, particularly with the numerical simulations, reinforcing the robustness of these model predictions.

The scaling relations of a range of galaxy properties and their evolution are also studied and compared to other work in the literature. For example, we present the relationships between halo mass, stellar mass, galaxy radius, cold gas phase metallicity, SFR, and rest-frame and observed luminosity (with and without dust attenuation). We present all of these relationships at $z=6$, $z=8$, and $z=10$, and provide extensive tables and fitting formulae describing the results. We hope that these predictions will be helpful both for interpreting future observations as well as for planning high-resolution zoom-in simulations.

By forward modeling the predicted star formation and chemical enrichment histories into the observed frame IR magnitude for \textit{JWST} NIRCam filters, we are able to select galaxies according to criteria intended to mimic representative wide, deep, and lensed \textit{JWST} surveys. We show the predicted distributions of physical properties for objects that would be detected in these surveys. We found that wide- and deep-field surveys will be able to probe galaxies down to $M_* \sim 10^{8}$\Msun\ and $\sim 10^{7}$\Msun\ at $z=4$, respectively. Our model also predicts that high-redshift galaxies are intrinsically brighter than their lower-redshift counterparts of similar mass. Thus, the $M_*$ corresponding to the detection limit evolves rather mildly as a function of redshift (see Table \ref{table:limit_prop}).

Taking advantage of our model's efficiency, we are able to quantify some of the effects of the uncertainties in the empirical physical recipes on the resultant galaxy population by systematically varying several key model parameters. We find that the gas depletion time and star formation relation have degenerate effects on the massive galaxy populations, while stellar feedback strongly influences star formation in low-mass halos. On the other hand, the current implementation of photoionization feedback from a metagalactic UV background, as well as AGN feedback, do not have a strong impact on our model predictions at these epochs.

We summarize our main conclusions below. 
\begin{enumerate}
    \item  The predictions from our fiducial model, which was calibrated only at $z=0$, are consistent with observed SMFs and SFRFs between $z \sim 4$--8. This suggests that our implementation of the physical processes that shape galaxy populations may still be reasonably accurate at these very early times.
    
    \item High-redshift SMFs from \textit{JWST} will help constrain our physical understanding of the buildup of galaxies. At high masses, the mass-function shape is significantly affected by the star-formation efficiency or timescale. At the low-mass end it is most influenced by the modeling of stellar-driven outflows. 

    \item We find that high-redshift galaxies are intrinsically brighter than their lower-redshift counterparts of similar stellar masses due to their higher SFR and younger stellar populations. In addition to that, massive galaxies at high redshifts have higher dust-extincted luminosities due to lower dust content. This results in a fixed \textit{observed-frame} $m_\text{F200W}$ selecting populations with similar stellar mass limits across the redshift range $z = 4$--10.

    \item We find that changing the slope of the scaling of the mass loading factor of stellar-driven outflows with galaxy circular velocity, $\alpha_\text{rh}$, changes both the slope of the low-mass end of the SMF and the mass where the SMF and LF cut off due to inefficient cooling. Changing $\alpha_\text{rh}$ from a value of 2.0 to 3.6 results in a change in the number density of the lowest mass galaxies ($M_* \lesssim 10^8$ \Msun) of 0.5--1 dex.  We find that changing the star formation efficiency by a factor of two has a negligible impact on the low mass end of the SMF, but changes the number density of massive galaxies by $\sim0.1$--$\sim 0.25$ dex.

    \item  There is remarkably good agreement between the predicted SMF from theoretical simulations based on different techniques and codes, in particular between our semi-analytic models and numerical hydrodynamical simulations. Larger discrepancies (up to nearly 1 dex for the lowest mass galaxies, and also for very massive galaxies) are seen between models when semi-empirical models are included in the comparison. Future observations with \textit{JWST} will be able to help discriminate between these models.

    \item Anticipated wide-field (deep-field) \textit{JWST} NIRCam surveys will be able to probe galaxies with rest-frame $M_\text{UV} \sim -17.28$ (-14.72) at $z\sim4$ and -18.73 (-16.06) at $z\sim10$, which corresponds to $\log M_*/$\Msun $\sim 7.96$ (6.78) at $z \sim 4$ and 7.51 (6.35) at $z \sim 10$ and $\log \text{SFR}/(\text{\Msun\ yr}^{-1}) \sim -0.4$ (-1.54) at $z\sim 4$ and -0.10 (-1.24) at $z\sim10$.
    
    \item None of models considered in this paper predict low-mass truncations or turnovers in the SMF at fluxes bright enough to be observed by \textit{JWST}, even in lensed fields. 
\end{enumerate}

\section*{Acknowledgements}

The authors of this work would like to thank St\'{e}phane Courteau et al. for organizing the `The Physics of Galaxy Scaling Relations and the Nature of Dark Matter' conference in Kingston, Ontario, which fostered significant discussions related to this work. The authors thank Aldo Rodr\'iguez-Puebla, Viraj Pandya, Rychard Bouwens, Takashi Okamoto, Eli Visbal, Peter Behroozi, and David Spergel for useful discussions. We also thank Sandro Tacchella and Christina Williams for providing unpublished results and their valued input. We also thank the anonymous referee for comments that improved this paper. AY and RSS thank the Downsbrough family for their generous support, and gratefully acknowledge funding from the Simons Foundation.

\bibliographystyle{mnras}
\bibliography{library.bib}

\appendix

\section{Tabulated values for selected distribution functions}
\label{appendix:a}
\setcounter{table}{0} \renewcommand{\thetable}{A\arabic{table}}

Tabulated SMFs and SFRFs from our fiducial model are provided in the tables \ref{table:tab_SMF}--\ref{table:tab_SFRF}. Other UV LFs shown in this work are available online.

\begin{table}
    \centering
    \caption{Tabulated SMFs at $z = 4$--10 from our fiducial model.}
    \label{table:tab_SMF}
    \scalebox{0.85}{
    \begin{tabular}{cccccccccc}
        \hline
        $\log{M_*}$ & \multicolumn{7}{c}{$\log_{10}(\phi\,[\text{dex}^{-1}\,\text{Mpc}^{-3}])$} \\
        $[\text{\Msun}]$ & $z = 4$ & $z = 5$  & $z = 6$  & $z = 7$  & $z = 8$  & $z = 9$  & $z = 10$   \\
        \hline
        4.0 & 0.06 & 0.21 & 0.32 & 0.42 & 0.31 & 0.19 & 0.07 \\
        4.5 & 0.22 & 0.37 & 0.44 & 0.33 & 0.19 & 0.01 & -0.23 \\
        5.0 & 0.20 & 0.24 & 0.21 & 0.06 & -0.09 & -0.36 & -0.71 \\
        5.5 & -0.00 & -0.05 & -0.11 & -0.26 & -0.49 & -0.73 & -1.05 \\
        6.0 & -0.22 & -0.29 & -0.44 & -0.62 & -0.87 & -1.16 & -1.55 \\
        6.5 & -0.51 & -0.58 & -0.76 & -1.02 & -1.27 & -1.61 & -2.04 \\
        7.0 & -0.80 & -0.96 & -1.13 & -1.39 & -1.69 & -2.04 & -2.55 \\
        7.5 & -1.12 & -1.30 & -1.52 & -1.81 & -2.19 & -2.60 & -3.20 \\
        8.0 & -1.44 & -1.66 & -1.97 & -2.28 & -2.69 & -3.27 & -3.89 \\
        8.5 & -1.82 & -2.07 & -2.35 & -2.80 & -3.28 & -3.93 & -4.70 \\
        9.0 & -2.16 & -2.46 & -2.84 & -3.36 & -3.97 & -4.69 & -5.57 \\
        9.5 & -2.53 & -2.90 & -3.39 & -4.01 & -4.78 & -5.71 & -6.76 \\
        10.0 & -2.94 & -3.45 & -4.05 & -4.86 & -5.78 & -6.86 & -8.23 \\
        10.5 & -3.42 & -4.08 & -5.02 & -6.12 & -7.23 & -8.99 & -10.87 \\
        11.0 & -4.54 & -5.88 & -7.14 & -8.78 & -11.04 & -99.99 & -99.99 \\
        \hline
    \end{tabular}
    }
\end{table}

\begin{table}
    \centering
    \caption{Tabulated SFRFs at $z = 4$--10 from our fiducial model.}
    \label{table:tab_SFRF}
    \scalebox{0.85}{
    \begin{tabular}{cccccccccc}
        \hline
        $\log{\text{SFR}}$ & \multicolumn{7}{c}{$\log_{10}(\phi\,[\text{dex}^{-1}\,\text{Mpc}^{-3}])$} \\
        $[\text{\Msun\ yr}^{-1}]$ & $z = 4$ & $z = 5$  & $z = 6$  & $z = 7$  & $z = 8$  & $z = 9$  & $z = 10$ \\
        \hline
        -5.0 & -0.00 & 0.05 & -0.06 & -0.05 & -0.14 & -0.21 & -0.35 \\
        -4.5 & 0.07 & 0.16 & 0.20 & 0.11 & 0.06 & -0.01 & -0.25 \\
        -4.0 & 0.18 & 0.36 & 0.38 & 0.40 & 0.33 & 0.21 & 0.07 \\
        -3.5 & 0.05 & 0.19 & 0.24 & 0.25 & 0.16 & 0.03 & -0.15 \\
        -3.0 & -0.15 & -0.05 & 0.01 & -0.06 & -0.12 & -0.33 & -0.57 \\
        -2.5 & -0.34 & -0.33 & -0.32 & -0.37 & -0.52 & -0.70 & -0.97 \\
        -2.0 & -0.64 & -0.63 & -0.61 & -0.76 & -0.87 & -1.12 & -1.40 \\
        -1.5 & -0.89 & -0.91 & -0.95 & -1.10 & -1.28 & -1.55 & -1.87 \\
        -1.0 & -1.21 & -1.26 & -1.30 & -1.51 & -1.70 & -2.00 & -2.40 \\
        -0.5 & -1.52 & -1.56 & -1.71 & -1.87 & -2.14 & -2.48 & -2.96 \\
        0.0 & -1.84 & -1.93 & -2.13 & -2.39 & -2.63 & -3.11 & -3.63 \\
        0.5 & -2.19 & -2.32 & -2.49 & -2.83 & -3.23 & -3.73 & -4.38 \\
        1.0 & -2.51 & -2.68 & -2.95 & -3.40 & -3.87 & -4.48 & -5.21 \\
        1.5 & -2.82 & -3.14 & -3.55 & -4.03 & -4.63 & -5.34 & -6.28 \\
        2.0 & -3.29 & -3.71 & -4.26 & -4.88 & -5.58 & -6.56 & -7.64 \\
        \hline
    \end{tabular}
    }
\end{table}

\section{Tabulated values for selected scaling relations}
\label{appendix:b}
\setcounter{table}{0} \renewcommand{\thetable}{B\arabic{table}}

The tabulated SHMR from our fiducial model is provided in Table \ref{table:tab_SHMR}. Other scaling relations shown in this work are available online.

\begin{table*}
    \centering
    \caption{Tabulated SHMR at $z = 4$--10 from our fiducial model. These are the median values as shown in fig. \ref{fig:SHMR}. The upper and lower limits give represent the value of the 84th and the 16th percentile.}
    \label{table:tab_SHMR}
    \scalebox{0.95}{
        \begin{tabular}{cccccccccc}
            \hline
            & \multicolumn{7}{c}{$\log_{10}(M_*/M_\text{h})$} \\
            $\log_{10}(M_\text{h}/\text{\Msun})$ & $z = 4$ & $z = 5$  & $z = 6$  & $z = 7$  & $z = 8$  & $z = 9$  & $z = 10$   \\
            \hline
            8.75 & -4.44$^{+0.43}_{-1.19}$ & -4.19$^{+0.28}_{-0.73}$ & -4.09$^{+0.26}_{-0.52}$ & -4.09$^{+0.29}_{-0.55}$ & -4.03$^{+0.27}_{-0.56}$ & -4.09$^{+0.35}_{-0.59}$ & -4.14$^{+0.36}_{-0.67}$ \\ [0.1in]
            9.00 & -4.02$^{+0.29}_{-0.67}$ & -3.91$^{+0.26}_{-0.52}$ & -3.88$^{+0.26}_{-0.46}$ & -3.82$^{+0.25}_{-0.55}$ & -3.82$^{+0.28}_{-0.55}$ & -3.85$^{+0.32}_{-0.58}$ & -3.86$^{+0.32}_{-0.68}$ \\ [0.1in]
            9.25 & -3.69$^{+0.22}_{-0.46}$ & -3.66$^{+0.25}_{-0.51}$ & -3.61$^{+0.22}_{-0.42}$ & -3.60$^{+0.25}_{-0.51}$ & -3.60$^{+0.28}_{-0.54}$ & -3.61$^{+0.29}_{-0.49}$ & -3.68$^{+0.35}_{-0.73}$ \\ [0.1in]
            9.50 & -3.47$^{+0.21}_{-0.42}$ & -3.40$^{+0.21}_{-0.40}$ & -3.37$^{+0.19}_{-0.42}$ & -3.44$^{+0.30}_{-0.49}$ & -3.45$^{+0.32}_{-0.55}$ & -3.45$^{+0.34}_{-0.57}$ & -3.43$^{+0.30}_{-0.58}$ \\ [0.1in]
            9.75 & -3.22$^{+0.20}_{-0.37}$ & -3.21$^{+0.22}_{-0.39}$ & -3.18$^{+0.22}_{-0.42}$ & -3.19$^{+0.25}_{-0.51}$ & -3.19$^{+0.29}_{-0.53}$ & -3.22$^{+0.31}_{-0.49}$ & -3.21$^{+0.31}_{-0.61}$ \\ [0.1in]
            10.00 & -2.99$^{+0.19}_{-0.39}$ & -2.98$^{+0.21}_{-0.40}$ & -2.99$^{+0.25}_{-0.43}$ & -3.00$^{+0.27}_{-0.43}$ & -2.98$^{+0.27}_{-0.47}$ & -3.00$^{+0.30}_{-0.56}$ & -3.04$^{+0.30}_{-0.58}$ \\ [0.1in]
            10.25 & -2.80$^{+0.19}_{-0.40}$ & -2.80$^{+0.21}_{-0.42}$ & -2.80$^{+0.25}_{-0.39}$ & -2.81$^{+0.28}_{-0.51}$ & -2.85$^{+0.35}_{-0.44}$ & -2.86$^{+0.33}_{-0.53}$ & -2.92$^{+0.36}_{-0.50}$ \\ [0.1in]
            10.50 & -2.60$^{+0.21}_{-0.34}$ & -2.62$^{+0.23}_{-0.42}$ & -2.61$^{+0.29}_{-0.41}$ & -2.62$^{+0.29}_{-0.44}$ & -2.63$^{+0.30}_{-0.47}$ & -2.70$^{+0.33}_{-0.45}$ & -2.78$^{+0.39}_{-0.54}$ \\ [0.1in]
            10.75 & -2.42$^{+0.25}_{-0.39}$ & -2.44$^{+0.28}_{-0.43}$ & -2.42$^{+0.28}_{-0.44}$ & -2.47$^{+0.32}_{-0.44}$ & -2.49$^{+0.32}_{-0.50}$ & -2.57$^{+0.35}_{-0.53}$ & -2.60$^{+0.37}_{-0.53}$ \\ [0.1in]
            11.00 & -2.20$^{+0.29}_{-0.39}$ & -2.24$^{+0.32}_{-0.46}$ & -2.22$^{+0.32}_{-0.49}$ & -2.28$^{+0.36}_{-0.43}$ & -2.32$^{+0.34}_{-0.55}$ & -2.46$^{+0.41}_{-0.48}$ & -2.49$^{+0.38}_{-0.58}$ \\ [0.1in]
            11.25 & -2.02$^{+0.31}_{-0.42}$ & -2.05$^{+0.30}_{-0.43}$ & -2.12$^{+0.34}_{-0.50}$ & -2.12$^{+0.33}_{-0.48}$ & -2.25$^{+0.36}_{-0.57}$ & -2.34$^{+0.40}_{-0.53}$ & -2.41$^{+0.39}_{-0.51}$ \\ [0.1in]
            11.50 & -1.84$^{+0.28}_{-0.52}$ & -1.94$^{+0.31}_{-0.48}$ & -1.98$^{+0.33}_{-0.49}$ & -2.03$^{+0.34}_{-0.51}$ & -2.13$^{+0.39}_{-0.54}$ & -2.26$^{+0.42}_{-0.47}$ & -2.32$^{+0.43}_{-0.56}$ \\ [0.1in]
            11.75 & -1.73$^{+0.31}_{-0.45}$ & -1.80$^{+0.33}_{-0.54}$ & -1.88$^{+0.29}_{-0.45}$ & -1.96$^{+0.34}_{-0.49}$ & -2.09$^{+0.39}_{-0.49}$ & -2.18$^{+0.41}_{-0.50}$ & -2.32$^{+0.39}_{-0.44}$ \\ [0.1in]
            12.00 & -1.64$^{+0.28}_{-0.50}$ & -1.79$^{+0.30}_{-0.43}$ & -1.92$^{+0.35}_{-0.38}$ & -2.01$^{+0.32}_{-0.47}$ & -2.11$^{+0.34}_{-0.46}$ & -2.19$^{+0.33}_{-0.40}$ & -2.37$^{+0.40}_{-0.46}$ \\ [0.1in]
            12.25 & -1.77$^{+0.27}_{-0.36}$ & -1.87$^{+0.28}_{-0.38}$ & -1.99$^{+0.31}_{-0.43}$ & -2.06$^{+0.31}_{-0.43}$ & -2.13$^{+0.30}_{-0.44}$ & -2.29$^{+0.37}_{-0.45}$ & -2.42$^{+0.34}_{-0.43}$ \\
            \hline
        \end{tabular}
    }
\end{table*}

\section{Fitting functions for selected scaling relations}
\label{appendix:c}
\setcounter{table}{0} \renewcommand{\thetable}{C\arabic{table}}
The fitting parameters for the medians of selected scaling relations presented in \ref{sec:scaling} are provided in table \ref{table:scaling_fit}. As noted in the text, the correlations break down in the bright, massive galaxies due to the effect of dust attenuation, which is not accounted for in the fitting. Bright galaxies with $m_\text{m200W} < 27$ and $M_\text{UV} < -24$ are excluded from our fits.  Galaxies with  $\log({M_*}/\text{\Msun}) < 6$ are also excluded as the scatter increases below this mass (see fig. \ref{fig:corner_z6_trimmed}, \ref{fig:corner_z8_trimmed}, and \ref{fig:corner_z10_trimmed}. In general, the residuals on these fits range from a few percent to $\sim 15\%$.
\begin{table*}
    \centering
    \caption{Fitting parameters for the medians of selected scaling relations. For each pair of $X$-$Y$ relation, the fitting parameters for $Y=Ax+B$ are given below.}
    \label{table:scaling_fit}
    \begin{tabular}{ccccccccccc}
        \hline
         & \multicolumn{2}{c}{$m_\text{F200W}$-$M_\text{UV}$} & \multicolumn{2}{c}{$m_\text{F200W}$-$M_*$} & \multicolumn{2}{c}{$m_\text{F200W}$-SFR} & \multicolumn{2}{c}{$M_\text{UV}$-$M_*$} & \multicolumn{2}{c}{$M_*$-SFR} \\ 
         $z$ & A & B & A & B & A & B & A & B & A & B \\ 
         \hline
         4 & 0.9048 & -42.9055 & -0.4033 & 19.4533 & -0.3746 & 9.9795 & -0.4420 & 0.2893 & 0.9692 & -8.3461 \\
         5 & 0.9161 & -43.9573 & -0.4000 & 19.4269 & -0.3829 & 10.5306 & -0.4233 & 0.3931 & 0.9828 & -8.2576 \\
         6 & 0.9139 & -44.1058 & -0.3991 & 19.3298 & -0.3828 & 10.6324 & -0.4244 & 0.2154 & 0.9679 & -7.9788 \\
         7 & 0.9088 & -44.0984 & -0.3988 & 19.2443 & -0.3823 & 10.6624 & -0.4266 & 0.0423 & 0.9472 & -7.7057 \\
         8 & 0.9200 & -44.5426 & -0.4058 & 19.3903 & -0.3950 & 11.0885 & -0.4279 & -0.1031 & 0.9532 & -7.6551 \\
         9 & 0.9274 & -44.8604 & -0.3983 & 19.0822 & -0.3919 & 10.9837 & -0.4242 & -0.1430 & 0.9543 & -7.6178 \\
         10 & 0.9120 & -44.4035 & -0.3880 & 18.6637 & -0.3856 & 10.7586 & -0.4208 & -0.1673 & 0.9513 & -7.5384 \\
        \hline
    \end{tabular}
\end{table*}

\section{Impact of adopting updated cosmological parameters: a comparison with previous work}
\label{appendix:d}
\setcounter{table}{0} \renewcommand{\thetable}{D\arabic{table}}
\setcounter{figure}{0} \renewcommand{\thefigure}{D\arabic{figure}}
During the course of recalibrating the model parameters, we found that changing from the WMAP cosmology to the Planck cosmology alone has some significant effects on our predictions for high-redshift galaxy populations. For each iteration of our model (e.g. \citetalias{Somerville2008}; \citetalias{Somerville2015}; \citetalias{Yung2019}), the parameters are tuned to match a subset of $z \sim 0$ observables. Although the specific observations used for calibration have changed throughout the years as the available observational constraints improved, the qualitative evolution of galaxy populations remained similar. However, as we move to $z \gtrsim 4$, the difference arising from the adopted cosmology becomes more noticable. Furthermore, different works have used various approximate fitting functions to estimate the abundance of dark matter halos as a function of mass and redshift. For example, \citetalias{Somerville2008} and \citetalias{Somerville2015} used the \citet*{Sheth2001} fitting function, while in this work we use the updated fitting function from \citet{Rodriguez-Puebla2016}. In Fig. \ref{fig:HMF_check_2} we show the effects of both fitting function choice and cosmology on the halo mass function at $z\sim 10$--0. These factors can lead to differences in halo abundance of a factor of two or more.

In this appendix, we focus on results at $z = 4$ from this work and compare them to results from \citetalias{Somerville2008} and \citetalias{Somerville2015}. The \citetalias{Somerville2008} models used the WMAP3 cosmology, and results from these models are shown in \citep{Porter2014}, the 
\citet{Lu2014a} SAM comparison and the \citet{Song2016} and \citet{Duncan2014} comparison to observations. The updated multiphase models presented in \citet{Somerville2015} incorporate the same gas-partitioning and \molh-based SF recipes used here, but adopted the WMAP7 cosmology \ref{table:tab_cosmo}.

We compare the SMFs from \citetalias{Somerville2008} and the non-multiphase version of \citetalias{Somerville2015}, and then the SMFs from the multiphase version of \citetalias{Somerville2015} and from this work. Our goal is to show the weak residual effects of the choice of cosmology on the redshift evolution of galaxy properties even \emph{after} model recalibration.

The Planck cosmology predicts higher abundances of halos of all masses and at all redshifts than WMAP, although not by the same factor at all redshifts and masses, as can be seen in fig. \ref{fig:HMF_check_2}. In \citetalias{Yung2019}, we describe how we recalibrated the parameters controlling the physical processes in the SAM to produce nearly identical results for our calibration quantities at $z\sim 0$ (Appendix C in \citetalias{Yung2019}). As shown in fig. \ref{fig:cosmos}, the updated model produces a very small excess of bright galaxies compared to the previously published \citepalias{Somerville2015} model. The residual difference at high redshift due to cosmology is also weakly dependent on the way that physical processes are modeled in the SAM, as can be seen from the models with different star formation recipes run within each of the two cosmologies.

\begin{table}
    \centering
    \caption{Summary of cosmological parameters used in previous models.}
    \label{table:tab_cosmo}
    \begin{tabular}{cccc}
        \hline
        &WMAP3 & WMAP5 & Planck 2015 \\
        &\citepalias{Somerville2008} & \citepalias{Somerville2015} & (This Work) \\
        \hline
        $\Omega_m$ & 0.2383 & 0.28 & 0.308 \\
        $\Omega_\Lambda$ & 0.7617 & 0.72 & 0.692 \\
        $H_0$ & 73.2 & 70.0 & 67.8 \\
        $f_b$ & 0.1746 & 0.1658 &  0.1578 \\
        $\sigma_8$ & 0.761 & 0.81 & 0.831 \\
        $n_s$ & 0.958 & 0.96 & 0.9665 \\
        \hline
    \end{tabular}
\end{table}

\begin{figure}
    \includegraphics[width=\columnwidth]{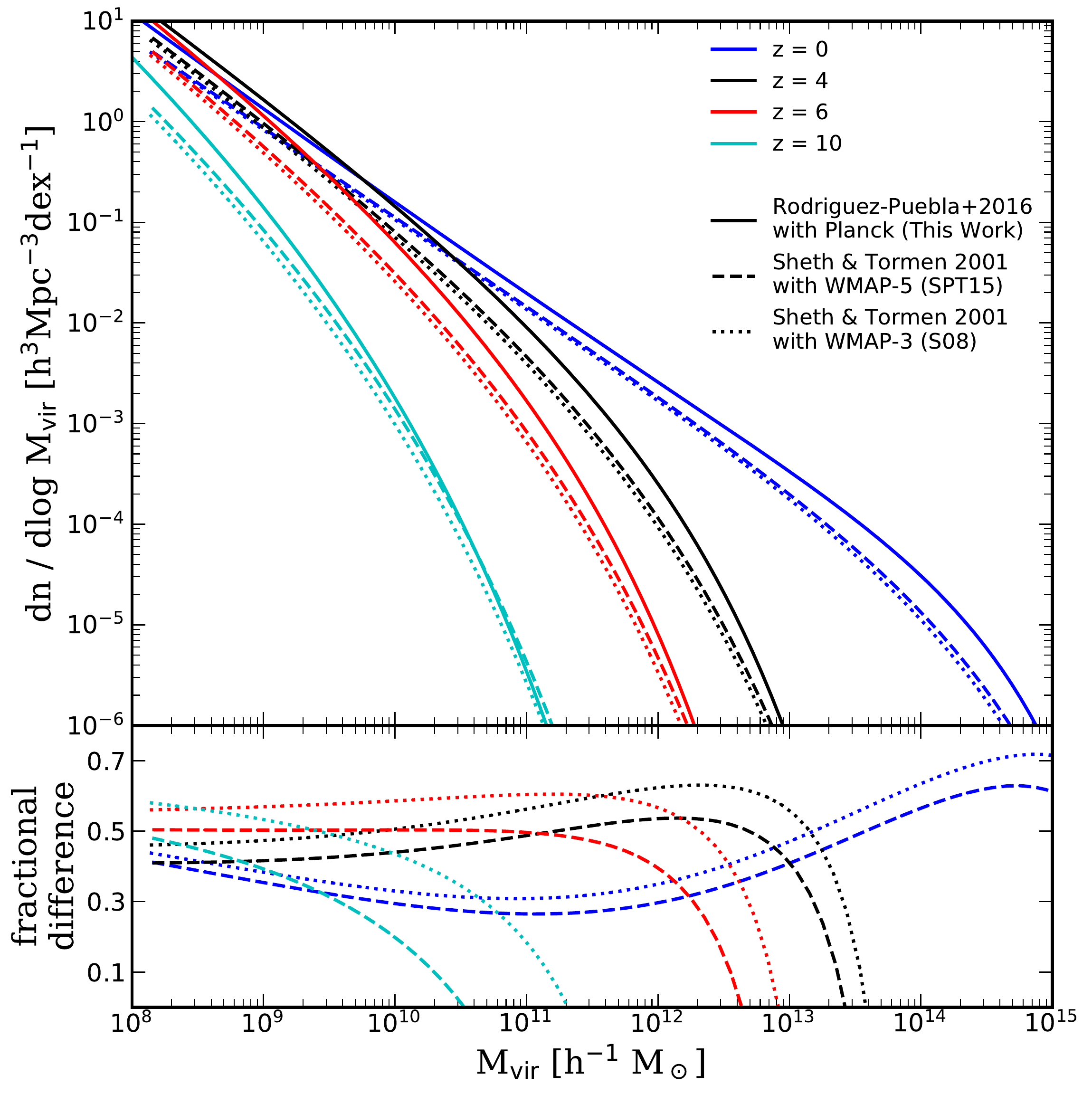}
    \caption{Halo mass functions from the fitting formulae of \citet{Sheth2001} and \citet{Rodriguez-Puebla2016} for different choices of cosmology: Planck (This Work, solid line), WMAP-7 \citepalias[dashed line]{Somerville2015}, WMAP-3\citepalias[dotted line]{Somerville2008}. The bottom panel shows the fractional difference relative to the RP16 Planck HMFs. The Planck cosmology generally predicts higher abundances of halos than WMAP cosmologies over a broad range of mass and redshift. }
    \label{fig:HMF_check_2}
\end{figure}

\begin{figure}
    \includegraphics[width=\columnwidth]{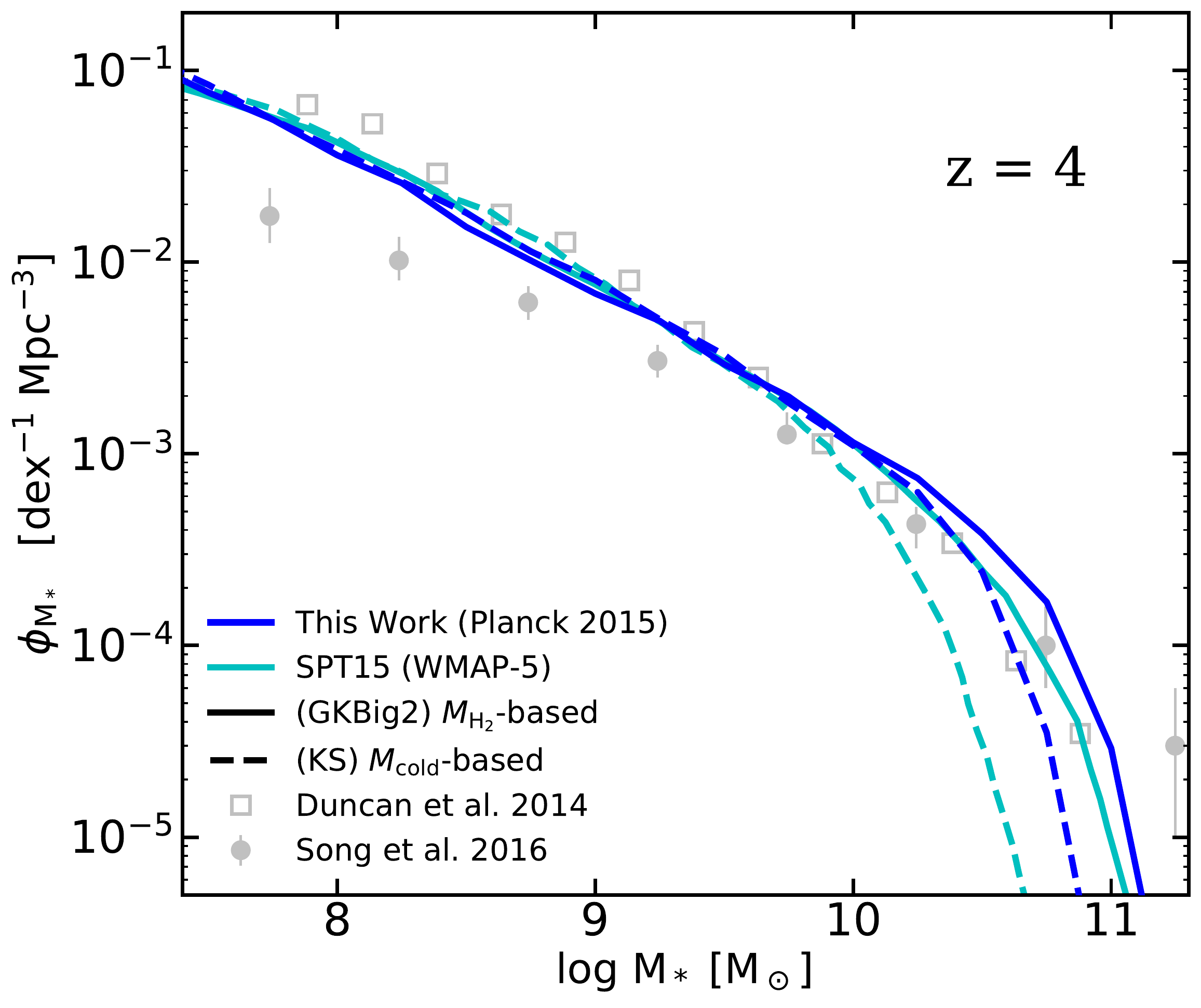}
    \caption{Stellar mass functions at $z = 4$ predicted by our SAM with different cosmologies and star formation recipes, \emph{after} retuning to match observational constraints at $z\sim 0$. These data are color-matched for the cosmology, where blue and cyan represent Planck 2015 and WMAP-5; and are style-matched for the star formation model, where solid and dashed lines represent the multiphase \molh-based GKBig2 and the non-multiphase $M_\text{cold}$-based KS model. This figure demonstrates that our predictions at high redshift show a weak dependence on cosmology after recalibration. }
    \label{fig:cosmos}
\end{figure}

\bsp
\label{lastpage}
\end{document}